\documentclass[journal]{IEEEtranTCOM}

\usepackage{times,amsmath,epsfig,latexsym,amssymb,psfrag}
\usepackage{cite,epstopdf}
\usepackage{algorithm}
 \usepackage{multirow}
\usepackage{algorithmic}
\usepackage{anyfontsize}
\usepackage{graphics, color, psfrag,amsfonts,mathtools}
\newtheorem{definition}{Definition}

\ifCLASSOPTIONtwocolumn
\skip\footins 0.9\baselineskip  plus 0.4\baselineskip  minus 0.2\baselineskip
\fi
\begin{document}
\ifCLASSOPTIONonecolumn
\title{\fontsize{23}{22}\selectfont  Joint Optimization of Throughput and Packet Drop Rate for Delay Sensitive Applications in TDD Satellite Network Coded Systems}
\else
\title{Joint Optimization of Throughput and Packet Drop Rate for Delay Sensitive Applications in TDD Satellite Network Coded Systems}
\fi
% \author{
%   \IEEEauthorblockN{Mohammad Esmaeilzadeh, Neda Aboutorab and Parastoo Sadeghi}
%}
 \author{
   \IEEEauthorblockN{Mohammad Esmaeilzadeh,~\IEEEmembership{Student Member,~IEEE,} Neda Aboutorab,~\IEEEmembership{Member,~IEEE,} and Parastoo Sadeghi,~\IEEEmembership{Senior Member,~IEEE}
\thanks{The authors are with the Research School of Information Sciences and Engineering, The Australian National University, Canberra, 0200, ACT, Australia
(e-mails: \{mohammad.esmaeilzadeh, neda.aboutorab, parastoo.sadeghi\}@anu.edu.au).}}% <-this % stops a space
\thanks{This work was supported under the Australian Research Council Discovery Projects and Linkage Projects funding schemes (project nos. DP120100160 and LP100100588).}
}
\maketitle

\ifCLASSOPTIONonecolumn
\vspace{-20mm}
\fi

\begin{abstract}
In this paper, we consider the issue of throughput and packet drop rate (PDR) optimization as two performance metrics for delay sensitive applications in network coded time division duplex (TDD) satellite systems with large round trip times (RTT). We adopt random linear network coding (RLNC) under two different scenarios, feedback-less and with feedback, and our goal is to jointly optimize the mean throughputs and PDRs of users in the system. For this purpose, we propose a systematic framework and start with formulating and optimizing these performance metrics for the single-user case. This framework enables us to analytically compare the performance metrics under different system parameters and settings. By comparing RLNC schemes under feedback-less and feedback scenarios for different RTTs, we show that the feedback-less schemes outperform the schemes with feedback in TDD systems with large RTTs.
Then, we extend the study of feedback-less RLNC schemes to the multi-user broadcast case. Here, we consider a number of different broadcast scenarios and optimize the system parameters such that the best overall performance is achieved. Furthermore, the complicated interplay of the mean throughputs and PDRs of different users with different packet erasure conditions in each of the considered broadcast scenarios is discussed.
\end{abstract}

\ifCLASSOPTIONonecolumn
\vspace{-5mm}
\fi
\ifCLASSOPTIONonecolumn
\vspace{-4mm}
\fi

\section{Introduction} \label{Section_1}

\ifCLASSOPTIONtwocolumn
\PARstart{D}{elivery}
\else
Delivery
\fi
of high data rate content with strict delay requirements is a constant challenge in many wireless communication systems. This is often due to multipath fading and shadowing effects, which eventually manifest themselves in the form of packet erasures. An example is live video broadcast to a group of wireless users in packet erasure channels. Due to different erasure events at different users, the sender is faced with various packet demands at any given time and hence choosing packets for transmission is not a trivial matter. This task becomes particularly difficult in time division duplex (TDD) systems with inherently large round trip times (RTT), such as satellite networks. In this case, the sender and receiver cannot transmit at the same time and providing feedback to the sender about the missing packets at each receiver can be extremely costly. Consequently, traditional automatic repeat request (ARQ) protocols are out of the question for such systems.

Instead, block-based transmission schemes such as raptor codes \cite{Shokrollahi:2006:IEEE-IT:Raptor} and random linear network coding (RLNC) \cite{Ho:IEEE-IT:2006:RLN} are more suitable for TDD systems with large RTT. RLNC refers to combining a block of $M$ packets using random coefficients from a finite field with diverse range of applications \cite{Chou:2003:Allerton:PNC, Lun:2008:Phy-Com:CRC,Eryilmaz:IEEE-IT:2008:DTG,Lucani:INFOCOM:2009:RLN, Lucani:NetCod:2009:BTD, Lucani:2009:GLOBECOM:RNC,Lucani:2010:ISIT:SysNC, Lucani:2012:IEEE-IT:NCD,Yazdi:2009:INFOCOM:ONC, Sorour:2009:PIMRC:NCARQ, Sorour:2010:ASMS:JCD, Swapna:2010:TDA, Nistor:2011:IEEE-JSAC:ODD, YANG:2012:ONLINE:ANC, Zeng:2012:ONLINE:JCD, Tran:2012:ASM}. In a \emph{rateless} broadcast scenario using RLNC, the sender then keeps transmitting different packet combinations until all receivers collect $M$ linearly independent coded packets and can hence obtain the block of $M$ original packets. This interval is referred to as the completion time. A lower completion time signifies a higher data throughput. However, in delay sensitive applications, the sender transmits a certain number of RLNC packets for any given block of packets and then moves onto a new block. Undecoded packets will have to be \emph{dropped} at the receivers upon the start of a new block. The following fundamental questions exist in TDD systems that use RLNC: Whether should the receivers provide feedback to the sender about their packet reception status? And if yes, how often? Also can the sender utilize this precious feedback to optimize the number of coded packet transmissions such that the average throughput is maximized or the packet drop rate (PDR) is minimized?

For TDD systems using RLNC transmission scheme, the authors in \cite{Lucani:INFOCOM:2009:RLN} considered optimization of the feedback frequency for a single-user case and then for the broadcast case \cite{Lucani:NetCod:2009:BTD}. Their objective was to minimize the completion time for delivering a block of $M$ packets to all users in a rateless fashion. That is, they did not consider a delivery deadline nor aimed to minimize the PDR. Instead, they assumed a priori that the transmission is broken into rounds. At the end of each round feedback should be provided from each user about the number of linearly independent coded packets still needed (referred to as the remaining degrees of freedom or rDOF). Based on this rDOF, the sender then decided about the number of coded packets to transmit in the next round before waiting to listen to another feedback. The number of coded packets for each rDOF were optimized such that the average completion time was minimized.

In a separate work, the authors in \cite{Sorour:2009:PIMRC:NCARQ, Sorour:2010:ASMS:JCD} considered the problem of PDR in TDD satellite systems using RLNC. However, given strict delay requirements of the considered application, they assumed a priori a feedback-less transmission scheme in \cite{Sorour:2010:ASMS:JCD} and aimed to meet a PDR threshold.

In this paper we consider a more general problem in two fronts. First, we wish to jointly maximize throughput and minimize the PDR within the delay requirements of the application and the imposed physical limitations of the channel such as RTT. This joint optimization is more desirable for realizing high data rate delay sensitive wireless applications. We note that there is a tradeoff between the two objectives. That is, for reducing PDR more coded packets of the same block should be sent so that each receiver is more likely to decode before the deadline. However, this results in increasing the completion time. So the optimum solution is not trivial, especially when multiple users have different PDR and throughput requirements or experience different packet erasure conditions. Second, for our joint optimization problem, we do not assume a priori whether feedback should be used or not in the system. Instead, we consider both feedback-less and feedback schemes in a unified systematic framework. Hence, comparison of the two schemes in terms of their throughput and PDR performance becomes possible under a variety of system parameters, such as the number of packets in a block, packet erasure probability, feedback erasure probability, RTT, delivery deadline, transmission rate, packet length, RLNC field size, etc. We also consider both normal RLNC and systematic RLNC (SRLNC)~\cite{Lucani:2010:ISIT:SysNC}, where in the latter all $M$ packets in a block are first broadcast uncoded before switching to normal RLNC transmission. Similar to the approach in \cite{Lucani:2012:IEEE-IT:NCD} and to gain insights into the system performance, we first solve the problem for the single-user case and then extend the analysis to broadcasting to multiple users.

To assess the performance of the proposed schemes, we compare our results with Round Robin (RR), as a simple scheduling scheme, and also with an ideal SRLNC scheme, where we assume that immediate feedbacks about the reception status of each user are available at the sender.

The main contributions and findings of this paper can be summarized as follows. First, we propose a systematic framework to analytically study RLNC for delay sensitive applications over TDD erasure channels. Then, employing the proposed framework, we formulate the mean throughput and PDR for RLNC and SRLNC schemes, and compare these schemes under feedback-less or feedback scenarios. Furthermore, the impact of different values of RTT and delivery deadline on the performance of these schemes is investigated. We observe that for the practical values of RTT and delivery deadline in satellite streaming applications, the feedback-less SRLNC scheme outperforms the other investigated schemes.
Second, we highlight the trade-off between the mean throughput and PDR and propose a joint optimization of these performance metrics for the feedback-less broadcasting schemes.
Furthermore, we present various system design approaches and demonstrate their effects on the performances of users with different packet erasure conditions. Finally, the optimum transmission schemes in terms of RLNC design parameters are obtained.

The rest of this paper is organized as follow. In the next section, we introduce our system model. Then in Section III, the formulation of throughput and PDR as the performance metrics for the single-user case is provided, and it is extended to broadcasting to multiple users in Section IV. Section V provides the numerical results. Finally, we conclude the paper in Section VI.

\section{System Model} \label{Section_2}

The system model consists of a satellite sender and a set of $N$ on-earth users. The sender is supposed to deliver a block of $M$ data packets, denoted by $\mathcal{M}=\{x_1, ..., x_M\}$, to the users before a specific delivery deadline $T_d$. We assume each data packet is independently useful to the users\footnote{Examples of this can be found in some video streaming protocols that employ multiple description coding~\cite{Goyal:2001:IEEE-SPM:MDC}, where receiving any subset of information is useful at the user. More examples of such systems are discussed in~\cite{Sadeghi:EURASIP:2010:OAN}.} and is composed of $n$ information bits. Moreover, we assume the transmission rate of the sender is $R$ bits per second (bps).

The channels between the sender and the users are assumed to be independent TDD channels (that is, nodes cannot transmit and receive at the same time) with packet error rates (PER) of $P_{e_i}$, $1\leq i \leq N$. In addition, it is assumed that channels are subject to large and equal RTTs, denoted by $T_{rt}$. This is a valid assumption in satellite communications as the distances between the satellite and the on-earth users are large and almost equal.

Moreover, the feedbacks, which are used to provide the sender with the reception status of packets at the users, are assumed to be composed of $n_{fb}$ bits and have erasure probability of $P_{e_{fb}}$.

\ifCLASSOPTIONonecolumn
\vspace{-3mm}
\fi
\subsection{Random Linear Network Coding (RLNC)}

Throughout this paper we employ two different types of RLNC, normal RLNC~\cite{Ho:IEEE-IT:2006:RLN} and \emph{systematic RLNC}~\cite{Keller:2008:NetCod:OBNC,Yazdi:2009:INFOCOM:ONC,Lucani:2010:ISIT:SysNC,Sorour:2010:ASMS:JCD, Sadeghi:2012:ONLINE:IDRL}, which are referred to as RLNC and SRLNC, respectively.

\subsubsection{Coding}

In the RLNC scheme, the sender only transmits coded packets, which are of the form $c_k=\sum_{m=1}^Ma_{mk}x_m$. The coding coefficients $a_{mk}$ are chosen randomly from a finite field of size $q$ ($\mathbb{F}_q$), and are sent along with packet's header and the linear combination of $M$ data packets. Therefore, a coded packet is composed of $l=h+n+Mg$ bits, where $h$ represents the number of bits allocated for the packet's header, $n$ is the number of bits for the linear combination of all $M$ data packets, and $g=\text{log}_2q$ is the number of bits used to represent the randomly chosen coding coefficients for each data packet.

In the SRLNC scheme, in an initial transmission phase, referred to as the \emph{systematic phase of transmission}, the original $M$ data packets are transmitted uncoded once. Then in the next phase of transmission, similar to RLNC, coded packets are transmitted to the users.

It should be noted that the uncoded packets are composed of $l_u=h+n$ bits. However, in order to have standard packet lengths in the system, we consider the length of both coded and uncoded packets to be $l$ bits and refer to them, in general, as \emph{RLNC packets} in the rest of this paper.

\subsubsection{Decoding} We start our explanation on the decoding with the following definition.
\begin{definition}
A received RLNC packet is said to provide one \emph{degree of freedom (DOF)} to a user, if it is linearly independent from previously received packets.
\end{definition}

A user, in order to decode the original $M$ data packets in a block, requires to collect $M$ DOF. However, this may require more than $M$ transmissions of uncoded/coded RLNC packets as the packets are subject to erasures and also it is probable that some received coded packets are not linearly independent from previously received ones.

\begin{definition}
For each user, we define the \emph{remaining DOF (rDOF)} at each time instance as the number of extra linearly independent packets that are needed by that user to be able to decode the entire set of $M$ data packets. It is clear that the rDOF is equal to $M$ at the beginning of the transmission, and rDOF of zero means that the block of $M$ data packets can be completely decoded.
\end{definition}

Having defined DOF and rDOF, it can be inferred that any successful reception of uncoded packets in the systematic phase of the SRLNC scheme will reduce the rDOF by one. However, for the coded packets, the rDOF reduction with each successful reception is not definite. The probability for the rDOF reduction upon successful reception of a coded packets is discussed next.

\subsubsection{Effect of Field Size $q$ on the rDOF Reduction} \label{Section_2_q}

In most of the existing works~\cite{Eryilmaz:IEEE-IT:2008:DTG,Keller:2008:NetCod:OBNC,Lucani:INFOCOM:2009:RLN, Lucani:NetCod:2009:BTD,Yazdi:2009:INFOCOM:ONC, Sorour:2009:PIMRC:NCARQ, Sorour:2010:ASMS:JCD, Swapna:2010:TDA, YANG:2012:ONLINE:ANC, Zeng:2012:ONLINE:JCD, Sadeghi:2012:ONLINE:IDRL, Tran:2012:ASM}, by assuming a very large field size $q$, the coded packets are considered to be always linearly independent. However, this simplifying assumption is not always practical due to the increased computational complexities of the decoding operations~\cite{Lucani:2012:IEEE-IT:NCD} and the amount of overhead imposed by large $q$. The authors in~\cite{Lucani:2009:GLOBECOM:RNC} were among the first to formulate the effect of field size $q$ in RLNC schemes by giving the probability of rDOF reduction in terms of the current rDOF and the size of $q$. This derivation on the effect of field size was used in~\cite{Lucani:2010:ISIT:SysNC, Lucani:2012:IEEE-IT:NCD, Nistor:2011:IEEE-JSAC:ODD}, and we utilize it here as well. By using the transition probability matrix introduced in~\cite{Lucani:2009:GLOBECOM:RNC}, we define $P_q^w(x,y)$ as the probability that $w$ successful received coded packets over $\mathbb{F}_q$ reduce the rDOF from $x$ to $y$.

\ifCLASSOPTIONtwocolumn
\begin{figure*}%[!t]
\centering
\includegraphics[width=6.5in]{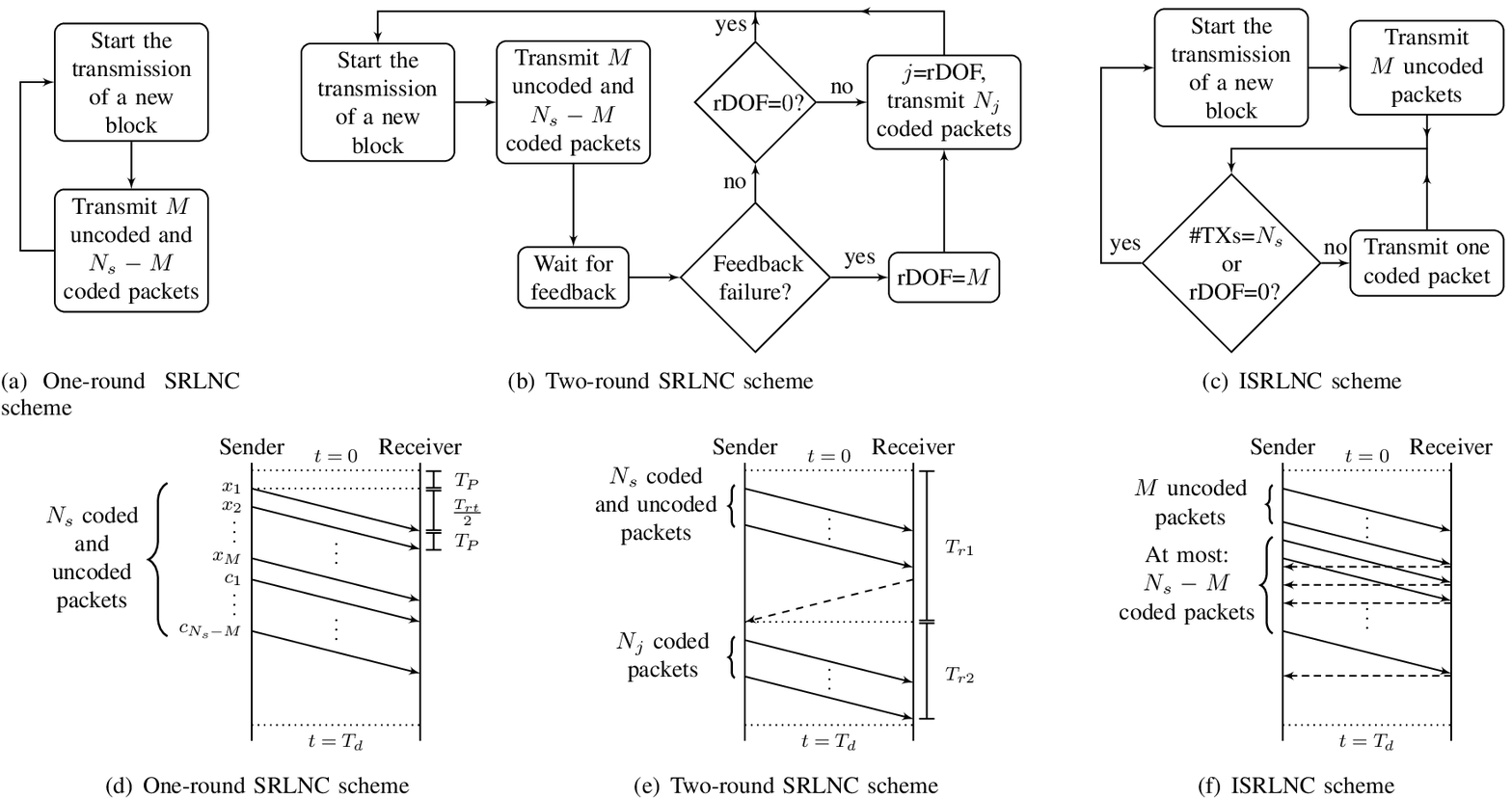}
\caption{Block diagrams and transmission timelines for the single-user case one- and two-round SRLNC schemes and also ISRLNC scheme (\#TXs stands for \emph{the total number of transmissions}). By considering all the $N_s$ packets in (a) and (d) to be either coded or uncoded, the diagrams for the one-round RLNC and RR schemes will be obtained, respectively. ISRLNC and RR are explained in Section~\ref{Section_5_a}.}  \label{Diagram1}
\end{figure*}
\fi

\subsection{Transmission Model}

Here, we explain the transmission model for the single-user case, and extend it to the multi-user case in Section~\ref{Section_4}. The channel is considered to be TDD with large RTT. The TDD nature of the channel forces the sender to stop its transmissions to be able to listen to the transmitted feedback from the user. Furthermore, large RTT makes the use of feedback to be extremely costly. As a result, it is not always beneficial for the sender to wait to listen to feedback after each single transmission. Hence similar to~\cite{Lucani:INFOCOM:2009:RLN}, we assume that the transmission is divided into a number of transmission rounds.
\begin{definition}
A \emph{round of transmission} is characterized by two stages, sending a number of RLNC packets back-to-back, and then waiting for feedback. The number of RLNC packets is predesigned for every rDOF value, and the feedback is then used to update the rDOF for the next round of transmission. It can be easily inferred that the duration of a round of transmission is lower bounded by $T_{rt}$.
\end{definition}

Unlike~\cite{Lucani:INFOCOM:2009:RLN}, where these transmission rounds were unlimitedly repeated until the rDOF becomes zero, in our model the existence of delivery deadline limits the number of transmission rounds. For instance, if we consider $T_{rt}$ to be equal to $250$ ms (i.e. a Geo-satellite system) and $T_d$ to be equal to $550$ ms (i.e. live video streaming applications), then it is clear that at most two transmission rounds are feasible.
This limited number of transmission rounds in our model causes another important distinction compared to the existing schemes with no delivery deadline~\cite{Lucani:INFOCOM:2009:RLN}, as in our model the feedback after the final round of transmission is \emph{not useful} and thus, not required. This can lead to a better performance by allowing the sender to either start the transmission of the next block of packets $\frac{T_{rt}}{2}$ seconds ahead in time or transmit more coded packets before the deadline.

In this paper, we study two transmission schemes,~\emph{one-round} and~\emph{two-round}, in a unified systematic framework\footnote{Although we consider up to two rounds of transmissions, our framework can be extended to consider more rounds.}. The one-round scheme, as shown in Fig.~\ref{Diagram1}(a), is in fact a feedback-less scheme, where for each block of $M$ packets, $N_s\geq M$ RLNC packets are transmitted back-to-back. Once the $N_s$ transmissions are completed, the transmission of the next block of packets is started. In the two-round scheme, as depicted in Fig.~\ref{Diagram1}(b), again $N_s\geq M$ RLNC packets are transmitted first, then the sender waits for the feedback from the user to update its rDOF, denoted by $j$. In case the feedback is not received, the sender assumes that the rDOF has remained unchanged, i.e. $j=M$. If $j=0$, the next block will be transmitted. Otherwise, $N_j\geq j$ coded packets will be transmitted back-to-back in the second round, and then the sender switches to the transmission of the next block of packets.

\ifCLASSOPTIONonecolumn
\begin{figure*}%[!t]
\centering
\includegraphics[width=6.5in]{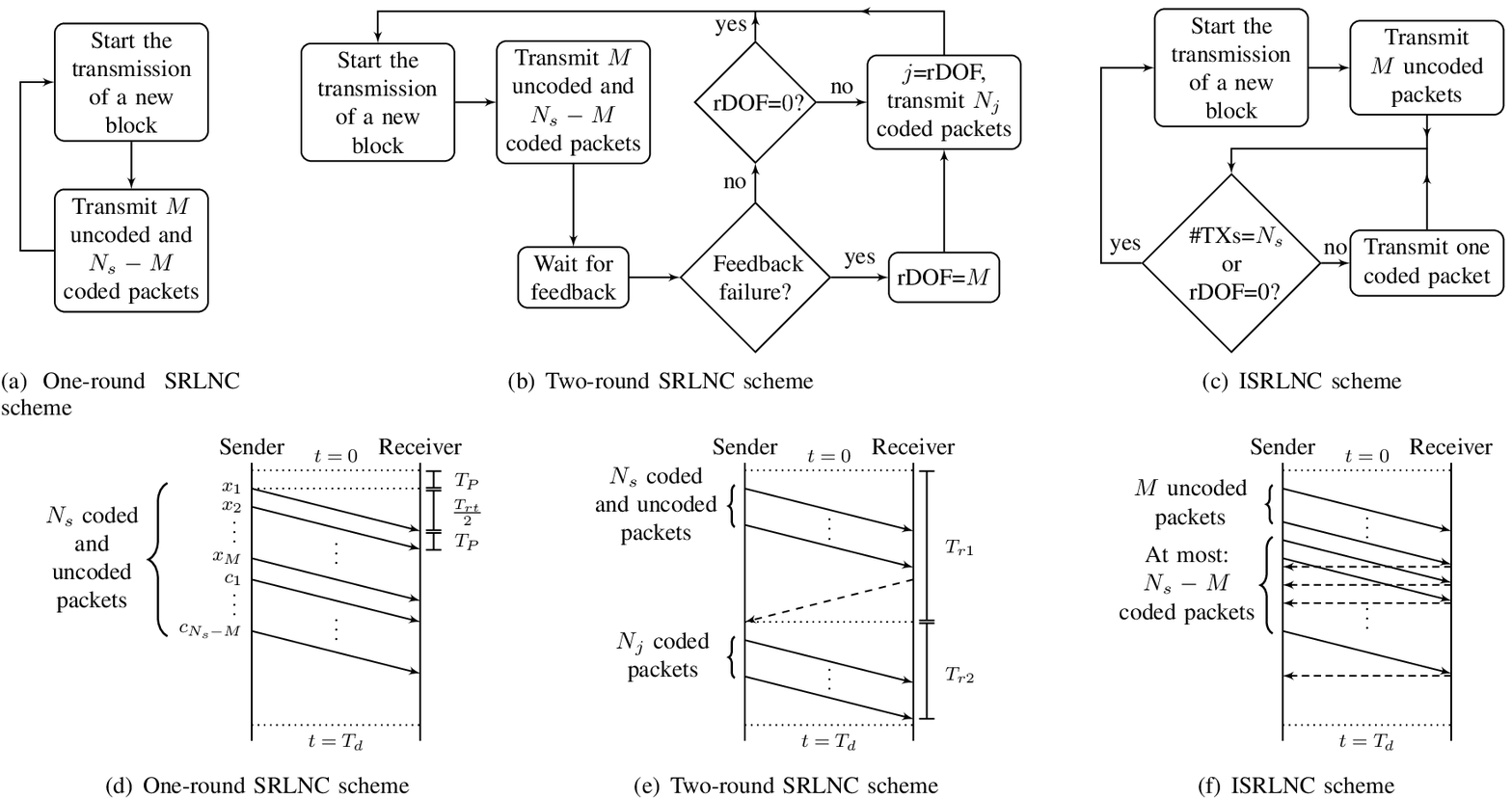}
\vspace{-5mm}
\caption{Block diagrams and transmission timelines for the single-user case one- and two-round SRLNC schemes and also ISRLNC scheme (\#TXs stands for \emph{the total number of transmissions}). By considering all the $N_s$ packets in (a) and (d) to be either coded or uncoded, the diagrams for the one-round RLNC and RR schemes will be obtained, respectively. ISRLNC and RR are explained in Section~\ref{Section_5_a}.}  \label{Diagram1}
\vspace{-8mm}
\end{figure*}
\fi

The one- and two-round schemes both need to satisfy the deadline requirement, i.e. the total transmission time, denoted by $T_{tot}$, should not exceed the deadline $T_d$, as illustrated in Figs.~\ref{Diagram1}(d) and~\ref{Diagram1}(e). Therefore, considering the propagation delay of $\frac{T_{rt}}{2}$ and the transmission times of an RLNC packet, $T_P=\frac{l}{R}$, and a feedback packet, $T_{fb}=\frac{n_{fb}}{R}$, we define feasible transmission schemes as follows:
\begin{definition} \label{Def_feasible_1}
A one-round RLNC or SRLNC transmission scheme with network coding (NC) parameters $M$, $N_s$ and $q$ is called \emph{feasible} when $N_s\geq M$ and $T_{tot}=N_{s}T_P+\frac{T_{rt}}{2}\leq T_d$.
\end{definition}

\begin{definition} \label{Def_feasible_2}
A two-round RLNC or SRLNC transmission scheme with NC parameters $M$, $N_s$, $N_j$ ($1\leq j\leq M$) and $q$ is called \emph{feasible} if $N_s\geq M$, $N_j\geq j$ and $T_{tot}=T_{r1}+T_{r2}\leq T_d$. Here, $T_{r1}$ and $T_{r2}$ represent the transmission times of the first and second rounds and are equal to $N_sT_P+T_{rt}+T_{fb}$ and $N_jT_P+\frac{T_{rt}}{2}$, respectively.
\end{definition}

\subsection{Performance Metrics}

Throughout this paper, we consider two performance metrics, mean throughput and PDR, which will be denoted by $\mathbb{E}\{\eta\}$\footnote{$\mathbb{E}\{~\}$ represents the expectation operator.} and $P_d$, respectively. For a user, we define $\mathbb{E}\{\eta\}$ as the expected number of data bits correctly decoded by that user per time unit before the deadline, and $P_d$ as the probability that a packet is not decoded by that user before the deadline.
In the next section, we start with formulating these two metrics for the single-user case, and then will extend it to the multi-user case in Section IV.

\section{Mean Throughput and PDR Formulation- Single-user Case} \label{Section_3}

In this section we focus on the single-user case.
Before discussing various schemes, we start with introducing some common notations and formulations, which will be used throughout this section.

We define $P(x,y,z)$ as the probability that transmission of $z$ coded packets over $\mathbb{F}_q$ reduces the rDOF from $x$ to $y$. Using $P_q^w(x,y)$ defined in Section~\ref{Section_2_q}, $P(x,y,z)$ can be expressed as
\begin{align}
P(x,y,z)=\sum_{w=x-y}^{z}{{z}\choose{w}}(1-P_e)^w(P_e)^{z-w}P_q^w(x,y)
\end{align}
where $P_e$ is the PER of the user.

For uncoded packets, we define $P_{sys}(M,m)$ as the probability of receiving $m$ uncoded data packets out of $M$ transmissions in the systematic phase of transmission as
\begin{align} \label{Eq_P_sys}
P_{sys}(M,m) = {{M}\choose{m}}(1-P_e)^mP_e^{M-m}
\end{align}
$P_{sys}(M,m)$ is in fact the probability that the rDOF reduces by $m$ after sending $M$ uncoded packets.

\subsection{One-round RLNC Scheme} \label{Section_3_a}

Considering the sender is transmitting a feasible number of coded packets $N_s$ for the transmission of a block of $M$ data packets, two different states are possible at the end of each round, \emph{success state} $\{S\}$ and \emph{failure state} $\{F\}$. If the transmission of $N_s$ coded packet brings down the user's rDOF from $M$ to zero (i.e. a successful/complete one-round transmission), the success state happens. Otherwise, the failure state happens (i.e. a failure/incomplete one-round transmission). The probabilities that each of these states occurs can be expressed as follows:
\begin{align}
P = \left\{
\begin{array}{l} \label{Eq_P_one_round_RLNC}
P_S=P(M,0,N_s) \\
P_F=1-P_S=1-P(M,0,N_s)
\end{array}\right.
\end{align}
and the corresponding throughput values will be
\begin{align} \label{Eq_eta_one_round_RLNC}
\eta = \left\{
\begin{array}{l}
\eta_S=\frac{Mn}{T_{tot}} \\
\eta_F=0
\end{array}\right.
\end{align}
Now using~\eqref{Eq_P_one_round_RLNC} and~\eqref{Eq_eta_one_round_RLNC}, $\mathbb{E}\{\eta\}$ and PDR can be defined as
\begin{align}
\mathbb{E}\{\eta\}=&\eta_SP_S+\eta_FP_F=\frac{Mn}{N_{s}T_P+T_{rt}/2}P(M,0,N_s) \label{Eq_RLNC_one_round_mean_eta}
\end{align}
\begin{align}
&P_d=P_F=1-P_S=1-P(M,0,N_s) \label{Eq_RLNC_one_round_mean_pdr}
\end{align}

From \eqref{Eq_RLNC_one_round_mean_eta} and \eqref{Eq_RLNC_one_round_mean_pdr}, the trade-off between the mean throughput and PDR, which is mostly affected by number of packets and transmissions ($M$ and $N_s$), can be concluded. For example to achieve a lower PDR for a constant $M$, sending more RLNC packets, i.e. larger $N_s$, is required. This improves $P_S$, but at the cost of reducing $\eta_S$, which may result in degradation of $\mathbb{E}\{\eta\}$.

Moreover, it could be inferred from \eqref{Eq_RLNC_one_round_mean_eta} and \eqref{Eq_RLNC_one_round_mean_pdr} that in the normal RLNC scheme, since the $M$ required DOF are not received at the user in state $\{F\}$, regardless of the value of rDOF, the whole block of $M$ data packets should be dropped without any contribution to the throughput. This motivates the idea of using SRLNC, where in addition to sending coded packets, the original data packets are sent uncoded in the initial phase of transmission. Since in Section~\ref{Section_2}  we have assumed  that the information in each of the $M$ packets in a block can be independently useful at the user, receiving even one of the $M$ uncoded packets has its own contribution to improving the throughput and PDR. This will be taken into account in the next subsection.

\subsection{One-round SRLNC Scheme} \label{Section_3_b}

\begin{figure*}%[!t]
\centering
\ifCLASSOPTIONonecolumn
\vspace{-1mm}
\fi
\includegraphics[width=6.6in]{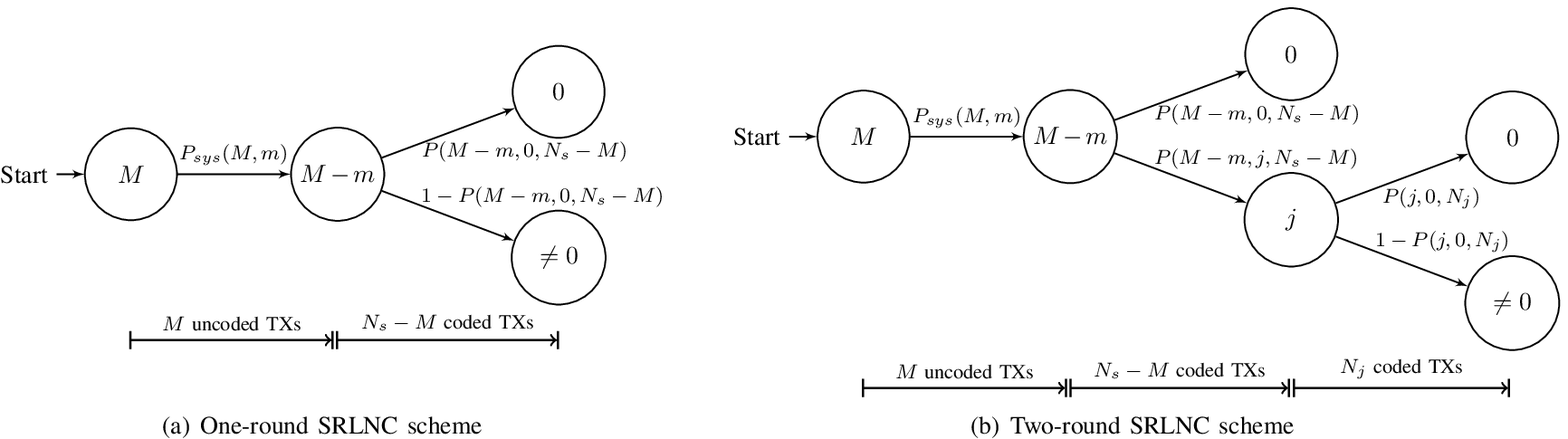}
\ifCLASSOPTIONonecolumn
\vspace{-4 mm}
\fi
\caption{State transition diagrams for SRLNC schemes  (TX: transmission). States represent the rDOF. We consider $P_{e_{fb}}=0$ in (b).}  \label{Diagram2}
\ifCLASSOPTIONonecolumn
\vspace{-8 mm}
\fi
\end{figure*}
For a feasible one-round SRLNC scheme, sending $N_s$ RLNC packets will again lead to either a complete ($\text{rDOF}=0$) or incomplete ($\text{rDOF}\neq0$) transmission, as shown in Fig.~\ref{Diagram2}(a). However, each of these two states can be further split into a number of sub-states, depending on the number of successfully received uncoded packets in the systematic phase of transmission. Assuming that $m$ uncoded packets are successfully received, the success and failure sub-states, denoted by $\{S|m\}$  ($0\leq m\leq M$) and $\{F|m\}$ ($0\leq m\leq M-1$), respectively, happen with the following probabilities and throughput values
\begin{align} \label{Eq_P_one_round_SRLNC}
P = \left\{
\begin{array}{l}
P_{S|m}=P(M-m,0,N_s-M) \\
P_{F|m}=1-P(M-m,0,N_s-M)
\end{array}\right.
\end{align}
\begin{align}
\eta = \left\{
\begin{array}{l}
\eta_{S|m}=\frac{Mn}{T_{tot}}\\
\eta_{F|m}=\frac{mn}{T_{tot}}
\end{array}\right.
\end{align}
Hence, the mean throughput for the one-round SRLNC scheme can be defined as
\begin{align}
\mathbb{E}\{\eta\}&=\sum_{m=0}^MP_{sys}(M,m)\Big[ \eta_{S|m}P_{S|m}+\eta_{F|m}P_{F|m}\Big]  \label{Eq_SRLNC_one_round_mean_eta}
\end{align}
%\begin{align}
%\text{E}\{\eta\}&=\sum_{m=0}^MP_{sys}(M,m)\Big[ \eta_{S|m}P_{S|m}+\eta_{F|m}P_{F|m}\Big]=  \label{Eq_SRLNC_one_round_mean_eta} \\
%&=\frac{Mn}{N_{s}T_P+T_{rt}/2}\sum_{m=0}^MP_{sys}(M,m)\Big[P(M-m,0,N_s-M)+m/M\left(1-P\left(M-m,0,N_s-M\right)\right)\Big] \nonumber
%\end{align}
where $P_{F|M}=0$ is used here. Moreover, the PDR can be written as
\begin{align}
P_d=\sum_{m=0}^{M-1}P_{sys}(M,m)P_{F|m}\frac{M-m}{M} \label{Eq_SRLNC_one_round_pdr}
\end{align}
%\begin{align}
%&P_d=\sum_{m=0}^MP_{sys}(M,m)P_{F|m}\frac{M-m}{M}=\sum_{m=0}^MP_{sys}(M,m)\left(1-P(M-m,0,N_s-M)\right)\frac{M-m}{M} \label{Eq_SRLNC_one_round_pdr}
%\end{align}

\ifCLASSOPTIONonecolumn
\vspace{-6 mm}
\fi
\subsection{Two-round RLNC Scheme} \label{Section_3_c}

Following a similar approach as in Sections \ref{Section_3_a} and \ref{Section_3_b}, a feasible two-round RLNC scheme might be completed after the first round, which is denoted by state $\{S\}$, or be completed or remain incomplete after the second round, which is denoted by states $\{j,S\}$ or $\{j,F\}$, respectively. Here, $j$, $1\leq j\leq M$, represents the rDOF after the first round of transmission. We also consider two additional states corresponding to the complete and incomplete transmissions following an undelivered feedback (feedback failure or FF for short). We denote these states by $\{FF,S\}$ and $\{FF,F\}$, respectively. Having defined all the possible states, we can derive their probabilities as follows:
\begin{align} \label{Eq_P_two_round_RLNC}
P = \left\{
\begin{array}{l}
P_S=(1-P_{e_{fb}})P(M,0,N_s) \\
P_{j,S}=(1-P_{e_{fb}})P(M,j,N_s)P(j,0,N_j) \\
P_{j,F}=(1-P_{e_{fb}})P(M,j,N_s)(1-P(j,0,N_j)) \\
P_{FF,S}=P_{e_{fb}}P(M,0,N_s+N_M) \\
P_{FF,F}=P_{e_{fb}}(1-P(M,0,N_s+N_M))
\end{array}\right.
\end{align}
For instance, $P_{FF,S}$ here represents the probability of a complete transmission following an undelivered feedback, which is obtained by multiplying $P_{e_{fb}}$ by the probability that rDOF reduces from $M$ to zero with total $N_s+N_M$ transmissions.
The throughput values of the states can then be written as
\begin{align} \label{Eq_eta_two_round_RLNC}
\eta = \left\{
\begin{array}{l}
\eta_S=\frac{Mn}{T_{r1}} \\
\eta_{j,S}=\eta_{FF,S}=\frac{Mn}{T_{tot}} \\
\eta_{j,F}=\eta_{FF,F}=0
\end{array}\right.
\end{align}
Thus, the mean throughput, $\mathbb{E}\{\eta\}$, and the probability of packets being dropped, $P_d$, can be obtained as
\begin{align}
\mathbb{E}\{\eta\}=\eta_SP_S+&\Big(\sum_{j=1}^{M}\eta_{j,S}P_{j,S}\Big)+\eta_{FF,S}P_{FF,S} \label{Eq_RLNC_two_round_mean_eta} \\[2mm]
P_d&=\Big(\sum_{j=1}^{M}P_{j,F}\Big)+P_{FF,F} \label{Eq_RLNC_two_round_pdr}
\end{align}

\ifCLASSOPTIONonecolumn
\vspace{-6 mm}
\fi
\subsection{Two-round SRLNC Scheme} \label{Section_3_d}

This transmission scheme is in fact a combination of the transmission schemes discussed in Sections \ref{Section_3_b} and \ref{Section_3_c} in the sense that it behaves similar to the one-round SRLNC in the first round and similar to the two-round RLNC, thereafter. Hence, for a feasible two-round SRLNC scheme, the possible states are similar to the ones defined in Section~\ref{Section_3_c}, except that we also take into account their dependence on the number of successfully received uncoded data packets $m$. Fig.~\ref{Diagram2}(b) shows a special case of this scheme with $P_{e_{fb}}=0$. Thus, the probabilities of the possible states can be expressed as
\ifCLASSOPTIONonecolumn
\begin{align}\label{Eq_P_two_round_SRLNC}
P = \left\{
\begin{array}{l}
P_{S|m}=(1-P_{e_{fb}})P(M-m,0,N_s-M) \\
P_{j,S|m}=(1-P_{e_{fb}})P(M-m,j,N_s-M)P(j,0,N_j) \\
P_{j,F|m}=(1-P_{e_{fb}})P(M-m,j,N_s-M)(1-P(j,0,N_j)) \\
P_{FF,S|m}=P_{e_{fb}}P(M-m,0,N_s+N_M-M) \\
P_{FF,F|m}=P_{e_{fb}}(1-P(M-m,0,N_s+N_M-M))
\end{array}\right.
\end{align}
\else
\begin{align}\label{Eq_P_two_round_SRLNC}
P = \left\{
\begin{array}{l}
P_{S|m}=(1-P_{e_{fb}})P(M-m,0,N_s-M) \\
P_{j,S|m}=(1-P_{e_{fb}})P(M-m,j,N_s-M)\\
~~~~~~~~~~~\times P(j,0,N_j) \\
P_{j,F|m}=(1-P_{e_{fb}})P(M-m,j,N_s-M)\\
~~~~~~~~~~~\times(1-P(j,0,N_j)) \\
P_{FF,S|m}=P_{e_{fb}}P(M-m,0,N_s+N_M-M) \\
P_{FF,F|m}=P_{e_{fb}}(1-P(M-m,0,N_s+N_M-M))
\end{array}\right.
\end{align}
\fi
As an example, $P_{j,F|m}$ here represents the conditional probability of an incomplete transmission with $j>0$ rDOF after the first round, given $m$ successfully received uncoded packets in the initial phase of transmission. Three terms are involved in this probability. The first term shows the probability of successfully receiving the feedback. The second term is the probability that the remaining $N_s-M$ transmissions in the first round can bring down the rDOF from $M-m$ to $j$, and the last term is the probability that $N_j$ transmissions in the second round cannot bring down the rDOF to zero.

The throughput values for the above-mentioned states can be written as
\begin{align}\label{Eq_eta_two_round_SRLNC}
\eta = \left\{
\begin{array}{l}
\eta_{S|m}=\frac{Mn}{T_{r1}}\\
\eta_{j,S|m}=\eta_{FF,S|m}=\frac{Mn}{T_{tot}}\\
\eta_{j,F|m}=\eta_{FF,F|m}=\frac{mn}{T_{tot}}\\
\end{array}\right.
\end{align}
From \eqref{Eq_P_two_round_SRLNC} and \eqref{Eq_eta_two_round_SRLNC}, $\mathbb{E}\{\eta\}$ and $P_d$ can be expressed as
\ifCLASSOPTIONonecolumn
\begin{align}
\mathbb{E}\{\eta\}=\sum_{m=0}^M P_{sys}(M,m) \bigg[&\eta_{S|m}P_{S|m}
+\Big(\sum_{j=1}^{M}\eta_{j,S|m}P_{j,S|m}+\eta_{j,F|m}P_{j,F|m}\Big)
\nonumber \\
&+\eta_{FF,S|m}P_{FF,S|m}+\eta_{FF,F|m}P_{FF,F|m} \bigg]  \label{Eq_SRLNC_two_round_mean_eta}\\
P_d=\sum_{m=0}^{M-1}P_{sys}(M,m)&\frac{M-m}{M} \bigg[ \Big( \sum_{j=1}^{M}P_{j,F|m} \Big) +P_{FF,F|m} \bigg]   \label{Eq_SRLNC_two_round_pdr}
\end{align}
\else
\begin{align}
\mathbb{E}\{\eta\}&=\sum_{m=0}^M P_{sys}(M,m) \bigg[\eta_{S|m}P_{S|m}
\nonumber \\
&+\Big(\sum_{j=1}^{M}\eta_{j,S|m}P_{j,S|m}+\eta_{j,F|m}P_{j,F|m}\Big)
\nonumber \\
&+\eta_{FF,S|m}P_{FF,S|m}+\eta_{FF,F|m}P_{FF,F|m} \bigg]  \label{Eq_SRLNC_two_round_mean_eta}\\
P_d=&\sum_{m=0}^{M-1}P_{sys}(M,m)\frac{M-m}{M} \bigg[ \Big( \sum_{j=1}^{M}P_{j,F|m} \Big) +P_{FF,F|m} \bigg]   \label{Eq_SRLNC_two_round_pdr}
\end{align}
\fi
It is worth noting that $P_{j,F|M}$ and $P_{FF,F|M}$ are assumed zero here, since these states do not happen.

\subsection{One-round versus Two-round Schemes} \label{Section_3_e}

Having obtained the performance metrics for the one-round and two-round schemes, now our goal is to find the best scheme among them for the values of $T_{rt}$ corresponding to satellite transmissions.

With a closer look at the derived equations for $\mathbb{E}\{\eta\}$ and $P_d$ in \eqref{Eq_P_one_round_RLNC} to \eqref{Eq_SRLNC_two_round_pdr}, it can be seen that a number of parameters play important roles in these equations. These parameters can be categorized into three groups: channel-enforced parameters, $P_e$, $P_{e_{fb}}$ and $T_{rt}$; application-enforced parameters, $n$, $h$, $n_{fb}$, $R$ and $T_d$; and NC design parameters, $q$, $M$, $N_s$ and $N_j$, $j=1,...,M$. All of these parameters are common between the one-round and two-round schemes, except for the number of transmissions in the second round $N_j$, which is not defined for the one-round schemes. Due to this difference between the NC design parameters, the comparison of these schemes becomes far from trivial.

Therefore, by considering both performance metrics concurrently, here we propose a framework that is capable of not only finding the best one-round and two-round transmission schemes, but also comparing them regardless of the discrepancy in the NC design parameters. While this general framework can be exploited to investigate the effect of any of the channel-enforced and application-enforced parameters, we only investigate the effect of RTT and the deadline, i.e $T_{rt}$ and $T_d$. The approach will be explained for the one-round and two-round RLNC schemes, but it is valid for the SRLNC schemes as well.

Considering Definitions~\ref{Def_feasible_1} and~\ref{Def_feasible_2}, the minimum delivery deadlines that the one-round and two-round RLNC schemes can meet for a fixed RTT are $T_{d1} = MT_P+\frac{T_{rt}}{2}$ and $T_{d2} = 2MT_P+3\frac{T_{rt}}{2}+T_{fb}$, respectively, 
%\begin{align}
%T_{d1} = MT_P&+T_{rt}/2 \\
%T_{d2} = 2MT_P&+3T_{rt}/2+T_{fb} \label{Eq_two_round_min_deadline}
%\end{align}
where for both schemes, we have set $N_s$ to be equal to $M$. Moreover, for the two-round scheme, we have considered the worst case scenario where the rDOF remains $M$ after the first round. Hence, the number of transmissions in the second round $N_M$ is also set to be equal to $M$.

It is clear that serving an application with the delivery deadline requirement of smaller than $T_{d1}$ is impossible. Also, for $T_{d1}\leq T_d<T_{d2}$, only one-round scheme is feasible. Therefore, in order to compare these two schemes, we focus on cases with $T_d\geq T_{d2}$, where both schemes are feasible.

In order to compare these schemes for such values of delivery deadline $T_d$, we jointly optimize the performance metrics, i.e. $\mathbb{E}\{\eta\}$ and $P_d$, for our proposed schemes. Therefore, this problem falls in the category of multi-objective (more specifically, bi-objective) optimization~\cite{Sawaragi:1985:TMO, Marler:2004:SMO, Ehrgott:2005:MCO}, and takes the following forms for the one- and two-round RLNC schemes, respectively:
\begin{align}
&\text{one-round scheme:~~~~}\max_{N_s}~[\mathbb{E}\{\eta\}, -P_d] \label{Eq_one_round_optimization}\\
&\text{two-round scheme:}\max_{N_1,...,N_M,N_s}~[\mathbb{E}\{\eta\}, -P_d]  \label{Eq_two_round_optimization}
\end{align}
It should be noted that $M$ and $q$ are also among the NC design parameters, but we consider them to have fixed values here and will discuss their effect on the system performance later in Section~\ref{Section_5_f}.

As mentioned earlier, the solution to the optimization problems in \eqref{Eq_one_round_optimization} and \eqref{Eq_two_round_optimization} can denote a feasible transmission scheme if the conditions in Definitions~\ref{Def_feasible_1} and~\ref{Def_feasible_2} are satisfied. Now, among all feasible transmission schemes, a solution for each of the problems in \eqref{Eq_one_round_optimization} and \eqref{Eq_two_round_optimization} is called \emph{optimal}, if it can maximize both of the objectives, i.e. $\mathbb{E}\{\eta\}$ and $-P_d$, simultaneously. However, as explained in Section~\ref{Section_3_a}, there is a trade-off between these two objectives and as a result, such optimal solutions do not exist for these problems.
Therefore, in order to solve the optimization problems, we propose to find \emph{Pareto optimal} solutions~\cite{Sawaragi:1985:TMO}.

\begin{definition} \label{Def_Pareto}
A feasible solution is said to be a \emph{Pareto optimal} solution, if no other feasible solution with larger $\mathbb{E}\{\eta\}$ and smaller $P_d$ exists. It is worth noting that a Pareto optimal solution forms one Pareto optimal point in the \emph{objective space}, i.e. the diagram of $\mathbb{E}\{\eta\}$ versus $P_d$.
\end{definition}

In fact, for comparing the one-round and two-round schemes, because a single optimal solution does not exist, we obtain and compare the Pareto optimal solutions.

There are various methods~\cite{Sawaragi:1985:TMO, Marler:2004:SMO, Ehrgott:2005:MCO} to obtain the Pareto optimal solutions. One simple but \emph{exhaustive} approach is to calculate $\mathbb{E}\{\eta\}$ and $P_d$ for all feasible values of $x$ and then select all the Pareto optimal solutions among them based on Definition~\ref{Def_Pareto}. This is possible for the one-round scheme in~\eqref{Eq_one_round_optimization}, since there is only one variable, $N_s$, with limited possible values\footnote{The conditions in Definition~\ref{Def_feasible_1} provide an upper- and a lower-bound for $N_s$.}. However, this approach may not be computationally efficient for the two-round scheme in~\eqref{Eq_two_round_optimization}, due to the number of variables.
Another possible approach is to combine the two objectives to relax the bi-objective problem into a single-objective problem. The most common technique to do this is the~\emph{weighted sum} method~\cite{Marler:2004:SMO}. By using this method, the problem in~\eqref{Eq_two_round_optimization} can be rewritten as
\begin{align} \label{Eq_weighted_sum_two_round}
\max_{N_1,...,N_M,N_s}~\{\lambda \mathbb{E}\{\eta\}-(1-\lambda)P_d\}
\end{align}

It is shown in~\cite{Ehrgott:2005:MCO} that solving the above problem for any desirable value of $\lambda$, $0\leq \lambda \leq 1$, will give a Pareto optimal solution. Therefore, we are able to obtain as many Pareto optimal solutions as required by solving \eqref{Eq_weighted_sum_two_round} for various values of $\lambda$. Details on solving~\eqref{Eq_weighted_sum_two_round} are provided in Appendix~\ref{Appendix_pareto}. We also show through an example how this weighted sum approach is computationally more efficient than the exhaustive approach.
Therefore, we employ the exhaustive and weighted sum methods to obtain the Pareto optimal solutions in the one-round and two-round schemes, respectively. Then we compare these schemes by observing the resulting Pareto optimal points in the objective space.

As will be shown later in Section~\ref{Section_5_c}, for satellite applications with large RTTs, which are the target of this paper, the one-round schemes result in superior performances compared to their two-round counterparts. Therefore, we focus on the one-round schemes for the rest of this paper.

\section{Mean Throughput and PDR Formulation- Multi-user Case} \label{Section_4}

In this section, we extend our study from the single-user case to a multi-user broadcast case. We consider $N$ users with independent erasure channels.  Here, for any set of parameters, $\mathbb{E}\{\eta\}$ and $P_d$ of each user for the one-round RLNC and SRLNC schemes can be computed by using \eqref{Eq_RLNC_one_round_mean_eta} and \eqref{Eq_RLNC_one_round_mean_pdr}, and \eqref{Eq_SRLNC_one_round_mean_eta} and \eqref{Eq_SRLNC_one_round_pdr}, respectively. We denote the performance metrics of the $i$-th user by $\mathbb{E}\{\eta_i\}$ and $P_{d_i}$.

Now, from a system design perspective, we are interested in finding an appropriate operating point for the entire system. As mentioned earlier, a single optimal solution that maximizes $\mathbb{E}\{\eta\}$ and at the same time minimizes $P_d$ does not exist. Furthermore, we will later show in Section~\ref{Section_5_c} that Pareto optimal solutions provide a trade-off, not only between the performance metrics of each user, but also among the performance metrics of all users. Therefore, to obtain one operating point, some constraints should be imposed on the performance metrics of the users. This is what we refer to as the required \emph{quality of service (QoS)}. Hence, having a predefined QoS requirement, we form the optimization problem and obtain the NC design parameters, $N_s$, $M$ and $q$, such that the system's operating point is optimized according to the required QoS.
We choose to put the constrains on the PDRs and opt to maximize the throughputs. Thus, the general form of this problem can be written as
\ifCLASSOPTIONonecolumn
\begin{align} \label{Eq_Broadcast_problem}
\max_{N_s, M, q}F(\mathbb{E}\{\eta_1\},\cdots,\mathbb{E}\{\eta_N\})
~\text{subject to}~G(P_{d_1},\cdots,P_{d_N})\leq P_{th}
\end{align}
\else
\begin{align} \label{Eq_Broadcast_problem}
&\max_{N_s, M, q}F(\mathbb{E}\{\eta_1\},\cdots,\mathbb{E}\{\eta_N\}) \nonumber\\
&\text{subject to}~G(P_{d_1},\cdots,P_{d_N})\leq P_{th}
\end{align}
\fi
$F(\cdot)$ and $G(\cdot)$ functions, along with some practical broadcasting scenarios are discussed in the following subsection.

\subsection{Broadcasting Scenarios with Various QoS Criteria} \label{Section_4_b}

\subsubsection{Scenario I, Maximizing $\mathbb{E}\{\eta\}$ of a Single User Subject to a Constraint on its PDR} \label{Section_4_b_1}

In this scenario, we investigate how designing the system based on the QoS requirement of a single user can affect the performance of the remaining users. Therefore, considering $k$ to be the index of the user of interest, $F(\cdot)$ and $G(\cdot)$ can be defined as
\begin{align}
F(\mathbb{E}&\{\eta_1\},\cdots,\mathbb{E}\{\eta_N\}) = \mathbb{E}\{\eta_k\} \label{Eq_Broadcast_scenarioI_1}\\
&G(P_{d_1},\cdots,P_{d_N}) = P_{d_{k}} \label{Eq_Broadcast_scenarioI_2}
\end{align}

\subsubsection{Scenario II, Maximizing the Mean of Users' $\mathbb{E}\{\eta\}$ Subject to Constraints on PDR of all Users} \label{Section_4_b_2}

In this scenario, it is required that PDRs of all users do not exceed a predefined threshold and at the same time the mean $\mathbb{E}\{\eta\}$ of users is maximized. It can be easily inferred that if the PDR constraint is satisfied for the user with the worst PER, it will be satisfied for rest of the users as well. Thus, we can define the $F(\cdot)$ and $G(\cdot)$ functions in~\eqref{Eq_Broadcast_problem} as
\begin{align}
&F(\mathbb{E}\{\eta_1\},\cdots,\mathbb{E}\{\eta_N\}) =\frac{1}{N}\sum_{i=1}^N\mathbb{E}\{\eta_i\} \label{Eq_Broadcast_scenarioII_1}\\
&G(P_{d_1},\cdots,P_{d_N}) = P_{d_{i^*}}~~;~~i^*=\arg \max_i{P_{e_i}} \label{Eq_Broadcast_scenarioII_2}
\end{align}

\subsubsection{Scenario III, Maximizing the Mean of Users' $\mathbb{E}\{\eta\}$ Subject to a Constraint on the Mean PDR} \label{Section_4_b_3}

In this scenario, it is required that the mean $\mathbb{E}\{\eta\}$ of users be maximized, while the mean PDR is lower than a predefined threshold. Therefore, $F(\cdot)$ can be similarly defined by~\eqref{Eq_Broadcast_scenarioII_1} and $G(\cdot)$ takes the form
\begin{align}
G(P_{d_1},\cdots,P_{d_N}) =\frac{1}{N}\sum_{i=1}^N P_{d_{i}} \label{Eq_Broadcast_scenarioIII}
\end{align}
\subsubsection{Scenario IV, Maximizing the Mean of Users' $\mathbb{E}\{\eta\}$ Subject to a Constraint on the Geometric Mean PDR} \label{Section_4_b_4}
As we will see in Section~\ref{Section_5}, the PDR of users with different PERs take values from a wide range (e.g. from $10^{-15}$ to nearly 1). Hence, the arithmetic mean, presented in~\eqref{Eq_Broadcast_scenarioIII}, has more tendency toward the higher values of PDR. To overcome this problem, we propose to use the geometric mean instead. Thus, the definition in~\eqref{Eq_Broadcast_scenarioII_1} is still valid for $F(\cdot)$, however, we define $G(\cdot)$ as
\begin{align}
&G(P_{d_1},\cdots,P_{d_N})=\prod_{i=1}^N(P_{d_i})^{1/N}  \label{Eq_Broadcast_scenarioIV}
\end{align}

\section{Numerical Results} \label{Section_5}

In this section, we present the numerical results. Results will be discussed in three main parts. First, we consider the single-user case and present our results on comparing one-round and two-round schemes for various $T_{rt}$ and $T_d$ values. Second, we focus on the one-round schemes for the single-user case with large RTTs. Finally, we consider broadcasting to multiple users and provide the results for the one-round schemes with large RTTs. In the second and third parts, we compare our results with two other schemes, namely round robin (RR)~\cite{Eryilmaz:IEEE-IT:2008:DTG} and idealistic SRLNC (ISRLNC). To show the merit of the proposed NC schemes, RR is chosen as a sample of traditional scheduling techniques without the need of feedback. Furthermore, as our benchmark for optimum achievable performance of the proposed SRLNC scheme, we have considered ISRLNC, where immediate feedback is assumed to be available at the sender about the reception status of the users. This is the best performance that can be achieved by the proposed SRLNC scheme. These two comparison schemes are described next in more detail.

\ifCLASSOPTIONonecolumn
\vspace{-2mm}
\fi

\subsection{Comparison Schemes} \label{Section_5_a}

\subsubsection{Round Robin (RR)}

In the RR scheme, the sender transmits each block of $M$ data packets exactly $K>0$ times, i.e. the total number of transmissions is $N_s=KM$. Then, it moves to the next block and repeats the same process. For this transmission scheme to be feasible, we consider the system constraint on the delivery deadline. Therefore, $T_{tot}=N_sT_{P_ u}+\frac{T_{rt}}{2}\leq T_d$ should be satisfied, where $T_{P_ u}=\frac{l_u}{R}$ is the transmission time of an uncoded packet.

Similar to one-round RLNC and SRLNC schemes, this scheme uses no feedback. Therefore, RR scheme will be similar for both single-user and multi-user cases. The formulations of the performance metrics for this scheme are provided in Appendix \ref{Appendix_RR}.

\subsubsection{Idealistic SRLNC (ISRLNC)} \label{Section_5_a_2}

In the ISRLNC scheme, as shown in Figs.~\ref{Diagram1}(c) and \ref{Diagram1}(f), we consider that immediate feedbacks about the rDOF of all users are available at the sender after each transmission. Also, to obtain the absolute best achievable performance, the effect of field size $q$ is ignored and it is assumed that any $M$ coded packets are sufficient to decode the whole block of $M$ packets. Furthermore, we ignore the extra required bits for the transmission of coding coefficients and assume $T_P=T_{P_ u}$.

The existence of feedbacks in this scheme makes it a bit different to the proposed one-round RLNC, SRLNC, as well as RR schemes. In those schemes, exactly $N_s$ transmissions were carried out before moving to the next block, whereas in ISRLNC, it is possible that the sender moves to the next block before completing exactly $N_s$ transmissions. In fact, the sender transmits $M$ uncoded packets and $N_s-M$ coded packets for a block of $M$ data packets, unless all users get $M$ DOF earlier. Similar to RLNC and SRLNC schemes, a feasible ISRLNC scheme should satisfy $N_s\geq M$ and $N_sT_{P_u}+\frac{T_{rt}}{2}\leq T_d$.

Details on the formulations of $\mathbb{E}\{\eta\}$ and $P_d$ for this scheme are provided in Appendix~\ref{Appendix_ISRLNC}.

\subsection{Parameters Values} \label{Section_5_b}

Throughout this section, we set the transmission rate of the sender $R=5$ Mbps, the number of information bits in a packet $n=10000$, the number of bits in a feedback $n_{fb}=100$ and the number of header bits $h=80$. Moreover, the RTT, delivery deadline, number of packets and field size are set to $T_{rt}=250$ ms, $T_d=450$ ms, $M=10$ and $q=2^{10}$, respectively, unless stated otherwise. Other parameters will be specified when required.

\subsection{One-round Schemes versus Two-round Schemes- Single-user Case} \label{Section_5_c}

In this section, we compare the performance of the proposed one-round and two-round schemes for the single-user case. We consider the packet and feedback erasure probabilities to be $P_e=0.1$ and $P_{e_{fb}}=0$. Moreover, we assume three RTT values of $10$, $50$ and $250$ ms, and compare the proposed schemes for various values of deadline $T_d$. For each set of $\{T_{rt},T_d\}$, all the Pareto optimal points for the one-round schemes, and $360$ Pareto optimal points for the two-round schemes (by solving~\eqref{Eq_weighted_sum_two_round} for $360$ different values of $\lambda$) are obtained. The results are shown in Fig.~\ref{Fig_one_vs_two_round}. The main observations can be summarized as follows:

\begin{figure*}[!t]
\centering
\includegraphics[width=6.2in]{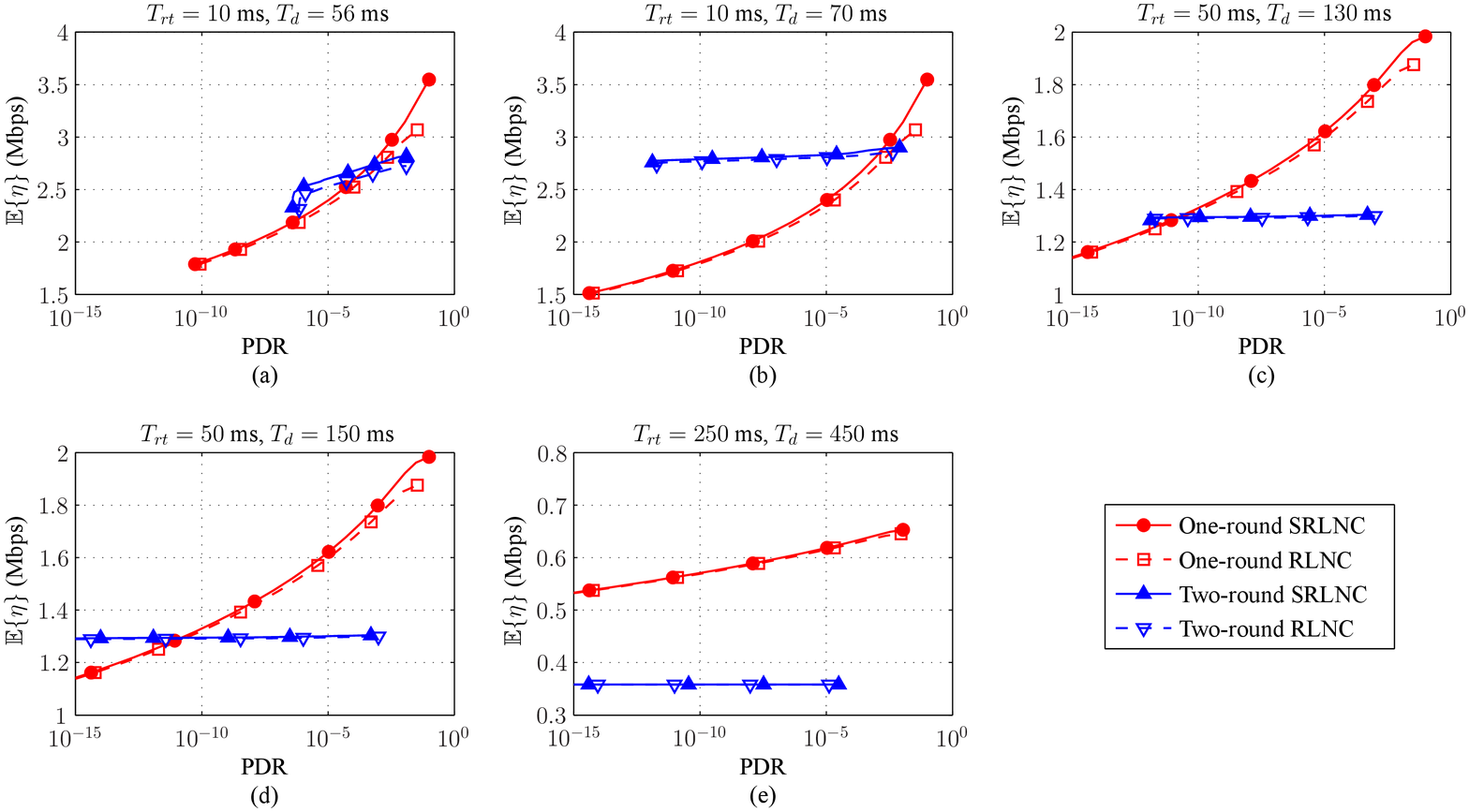}
\ifCLASSOPTIONonecolumn
\vspace{-5mm}
\fi
\caption{Mean throughput versus PDR for the one- and two-round RLNC and SRLNC schemes. Each point corresponds to a set of design parameter(s), $\{N_s\}$ for the one-round and $\{N_j,N_s\}$ for the two-round schemes.}  \label{Fig_one_vs_two_round}
\ifCLASSOPTIONonecolumn
\vspace{-7mm}
\fi
\end{figure*}
\begin{itemize}
\item By comparing Figs.~\ref{Fig_one_vs_two_round}(a) with~\ref{Fig_one_vs_two_round}(b) and also Figs.~\ref{Fig_one_vs_two_round}(c) with~\ref{Fig_one_vs_two_round}(d), we see that increasing $T_d$ allows larger number of transmissions, and as a result lower PDRs could be achieved.
\item By comparing cases with different RTTs, it is shown that, as expected, the mean throughput is inversely affected by the value of RTT, $T_{rt}$.
\item For $T_{rt}=10$ ms, the value of $T_d=T_{d2}=56$ ms in Fig.~\ref{Fig_one_vs_two_round}(a) is the minimum possible deadline that the two-round scheme can meet (in the worst case scenario). As the results suggest, based on the required PDR or $\mathbb{E}\{\eta\}$, either of the one- or two-round schemes could outperform the other one. By considering larger deadlines $T_d$ as in Fig.~\ref{Fig_one_vs_two_round}(b), two-round schemes are almost always superior to the one-round schemes.
\item For $T_{rt}=50$ ms, except for very small required PDR in Fig.~\ref{Fig_one_vs_two_round}(d), one-round schemes work better than the two-round counterparts.
\item In Fig.~\ref{Fig_one_vs_two_round}(e), for $T_{rt}=250$ ms, it can be seen that even for the relatively large delivery deadline of $T_d=450$ ms, still one-round schemes have superior performances over two-round schemes.
\item In Figs.~\ref{Fig_one_vs_two_round}(c) and~\ref{Fig_one_vs_two_round}(e), we chose $T_d$ in a way that the total allowed number of transmissions in the two-round schemes will be equal to those in Figs.~\ref{Fig_one_vs_two_round}(b) and~\ref{Fig_one_vs_two_round}(d), respectively. The observed outcome is that the minimum achievable PDR in Figs.~\ref{Fig_one_vs_two_round}(b) and~\ref{Fig_one_vs_two_round}(c), and also in Figs.~\ref{Fig_one_vs_two_round}(d) and~\ref{Fig_one_vs_two_round}(e) are equal for the two-round schemes.
\item In all cases, we observed that the SRLNC schemes are always superior to their RLNC counterparts.
\end{itemize}

Note that for the results presented in Fig.~\ref{Fig_one_vs_two_round}, we have considered $M$, $q$, $R$, $P_e$, $P_{e_{fb}}$, $n$ and $h$ to be fixed. Therefore, we cannot indicate which of the one-round or the two-round schemes results in a better performance in general. However, under the considered parameters, we can give a rule of thumb that if $T_{rt}\leq MT_P$ and $T_d$ allows around $2M\sim3M$ transmissions in the two-round schemes, then they work better than the one-round ones. However, when $T_{rt}$ is relatively large compared to $MT_P$, then the one-round schemes will outperform the two-round ones, especially for practical values of PDR, ranging from $10^{-3}$ to $10^{-6}$. For instance, in Fig.~\ref{Fig_one_vs_two_round}(e), with $T_{rt}=250$ ms and $T_P=2$ ms, to deliver $M=10$ data packets, the one-round schemes can blindly send $125$ more coded packets than the two-round ones, which clarifies why they result in far better performances.

\subsection{One-round RLNC and SRLNC Schemes- Single-user Case} \label{Section_5_d}

\begin{figure*}%[!t]
\centering
\includegraphics[width=6.1in]{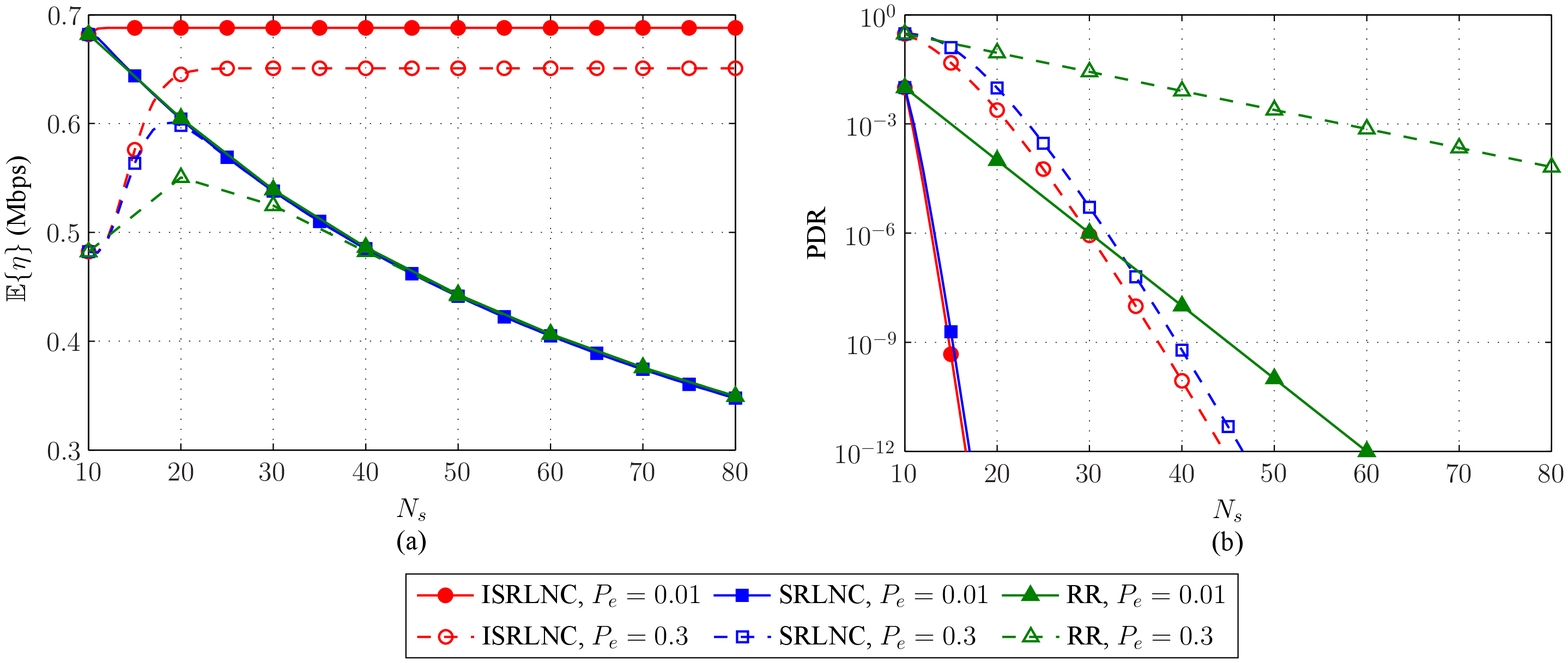}
\ifCLASSOPTIONonecolumn
\vspace{-4mm}
\fi
\caption{(a) Mean throughput and (b) PDR versus $N_s$ for two different values of PER}  \label{Fig_Eta_and_PDR_vs_Ns}
\ifCLASSOPTIONonecolumn
\vspace{-6mm}
\fi
\end{figure*}

In this section, we provide further results on the performance of the proposed one-round transmission schemes described in Sections~\ref{Section_3_a} and~\ref{Section_3_b}, and compare them with the performance of RR and ISRLNC schemes described in Section~\ref{Section_5_a} and Appendices~\ref{Appendix_RR} and~\ref{Appendix_ISRLNC}. Using the parameters specified in Section~\ref{Section_5_b}, Fig.~\ref{Fig_Eta_vs_PDR} depicts the mean throughput versus PDR for SRLNC and RLNC schemes by showing all the feasible points in the objective space. Furthermore, Fig.~\ref{Fig_Eta_and_PDR_vs_Ns} illustrates the mean throughput and the PDR versus $N_s$ for SRLNC, RR and ISRLNC schemes. Our observations are as follows:

\begin{figure}%[!t]
\centering
\includegraphics[width=3in]{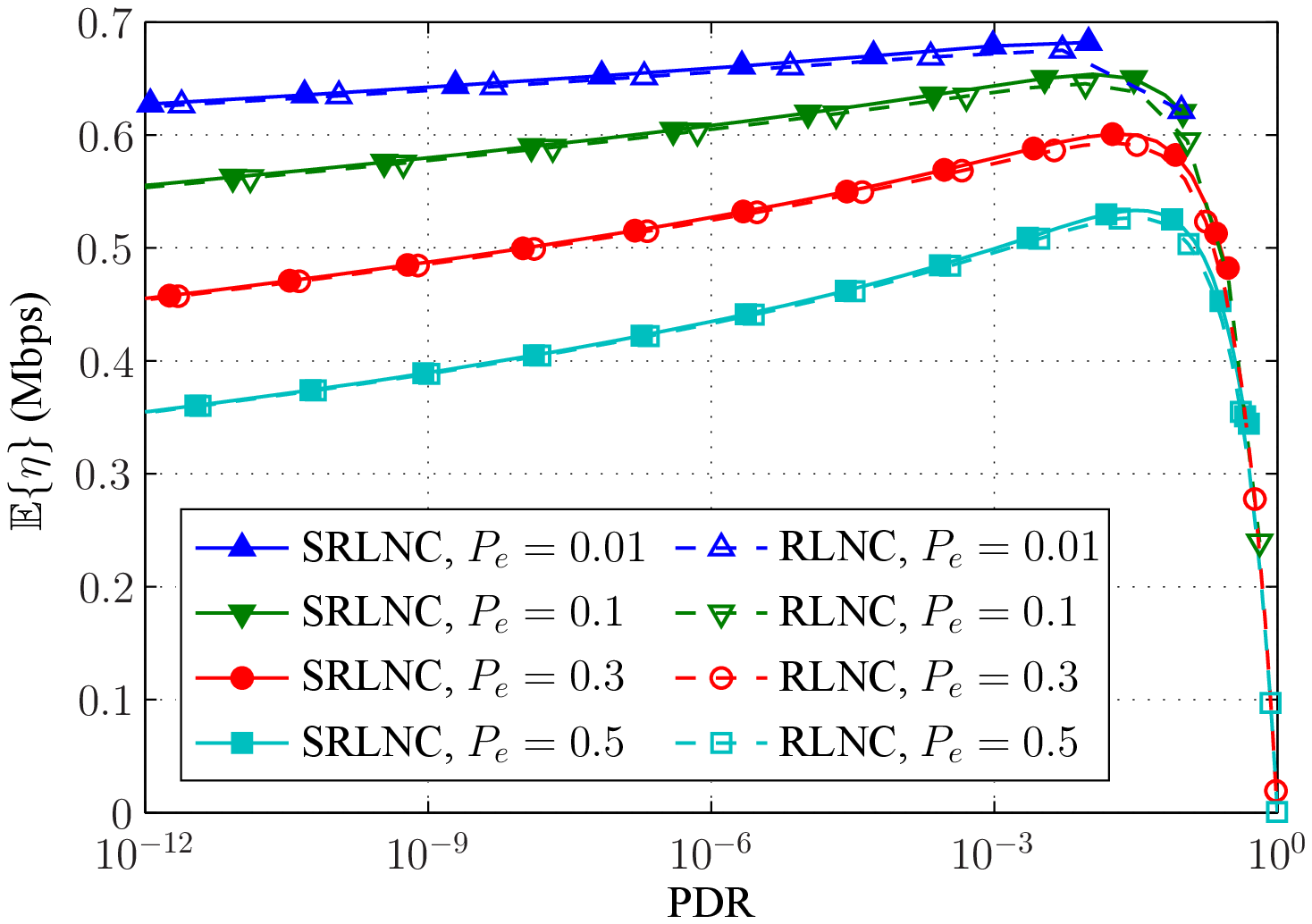}
\ifCLASSOPTIONonecolumn
\vspace{-5mm}
\fi
\caption{Mean throughput versus PDR for RLNC and SRLNC schemes for four different values of PER} \label{Fig_Eta_vs_PDR}
\ifCLASSOPTIONonecolumn
\vspace{-5mm}
\fi\end{figure}

\begin{itemize}
\item Similar to Fig.~\ref{Fig_one_vs_two_round}, Fig.~\ref{Fig_Eta_vs_PDR} reveals that SRLNC scheme provides a better performance compared to RLNC scheme. Thus, in the rest of this section, we will mainly focus on the SRLNC scheme.
\item In Fig.~\ref{Fig_Eta_and_PDR_vs_Ns}(a), depending on the value of PER, increasing the number of transmissions $N_s$ can result in either one or two phases in the mean throughput performance. For $P_e=0.01$, there exists only one phase, where SRLNC and RR performances decrease steadily, and that of  ISRLNC remains constant. In fact, because of the small PER, $N_s=10$ is enough for delivering all $M=10$ data packets with high probability, and increasing the value of $N_s$ has no advantage in terms of the mean throughput. For $P_e=0.3$, in an initial phase until a peak is reached, the mean throughput increases for all SRLNC, RR, and ISRLNC schemes. After that, further increase of $N_s$ leads to a similar result as for cases with $P_e=0.01$.
\item Considering the achieved mean throughput in Fig.~\ref{Fig_Eta_and_PDR_vs_Ns}(a), for large values of $N_s$, the SRLNC graphs converge to the common value of $\frac{Mn}{N_{s}T_P+T_{rt}/2}$. In fact, by increasing the value of $N_s$, the probability of successfully decoding $M$ data packets approaches one and thus the mean throughput is only affected by $N_s$. For the RR scheme, the common value $\frac{Mn}{N_{s}T_{P_u}+T_{rt}/2}$ is slightly higher, as $T_{P_ u}<T_P$ is used instead of $T_P$. For the ISRLNC scheme, the mean throughput never degrades with increasing the number of transmissions $N_s$. This is the result of having immediate feedbacks available at the sender, which prevents the sender from any unnecessary transmission.
\item The graphs in Fig.~\ref{Fig_Eta_and_PDR_vs_Ns}(b) show that the PDRs improve as $N_s$ increases. It is also observed that for all the studied schemes, PDRs are upper-bounded by the PERs. These maximum PDR values occur at $N_s=M$, and can be easily calculated by using~\eqref{Eq_SRLNC_one_round_pdr},~\eqref{Eq_RR_PDR} or~\eqref{Eq_ISRLNC_PDR}.
\item Results in Figs.~\ref{Fig_Eta_vs_PDR} and~\ref{Fig_Eta_and_PDR_vs_Ns} also show the trade-off between $\mathbb{E}\{\eta\}$ and $P_d$ for users with different PERs. Joint interpretation of the results in Figs.~\ref{Fig_Eta_and_PDR_vs_Ns}(a) and \ref{Fig_Eta_and_PDR_vs_Ns}(b) reveals the impact of $N_s$ on the performance of users with different PERs. This will be discussed in more detail in the next subsections.
\end{itemize}

\subsection{One-round SRLNC Scheme- Multi-user Case} \label{Section_5_e}

In this section, we discuss the broadcasting of a block of $M=10$ data packets to $N$ users employing SRLNC scheme with fixed field size $q=2^{10}$, through the scenarios described in Section~\ref{Section_4_b}. We assume four different classes of users, where each class has one of the fixed PER values of $P_e=[0.01, 0.1, 0.3, 0.5]$. We further consider that the ratios of the number of users belonging to these four values of PER (i.e. four classes) divided by the total number of users are equal to $\bar{N}=[0.3, 0.4, 0.2, 0.1]$, respectively. It is worth noting that knowing $\bar{N}$ is in fact enough for the calculation of $F(\cdot)$ and $G(\cdot)$ in all the considered scenarios for the SRLNC scheme, and also RR scheme, as the performance metrics for each user are independent of $N$. However for the ISRLNC scheme, as shown in the Appendix \ref{Appendix_ISRLNC}, $N$ is involved in the calculations of performance metrics of each user. Thus in Section~\ref{Section_5_g}, when comparing the results with the ISRLNC scheme, $N$ needs to be specified too.

To solve~\eqref{Eq_Broadcast_problem}, a threshold on $G(\cdot)$ is required. We choose two values of $10^{-3}$ and $10^{-6}$ for the threshold $P_{th}$ and find the optimum values of $N_s$ that maximize $F(\cdot)$ in any of the four scenarios. In Scenario I, the \emph{user of interest} is considered to be among the users in the class with $P_e=0.01$.

The results are presented in Table~\ref{Table_Ns_eta_PDR}, where in addition to the optimum values of $N_s$, the mean $\mathbb{E}\{\eta\}$ (denoted by $\overline{\mathbb{E}\{\eta\}}$ and obtained by~\eqref{Eq_Broadcast_scenarioII_1}) and mean PDR (denoted by $\overline{P_d}$ and obtained by~\eqref{Eq_Broadcast_scenarioIII} and~\eqref{Eq_Broadcast_scenarioIV}) are also shown. Comparing the results of these scenarios and using Figs.~\ref{Fig_Eta_and_PDR_vs_Ns}(a) and \ref{Fig_Eta_and_PDR_vs_Ns}(b), we can conclude that Scenarios I and II are in fact two extreme cases, as the performances in Scenario I are mostly in the favor of users with low PER and those in Scenario II are mostly in the favor of users with high PERs. For instance, in Scenario I with $P_{th}=10^{-6}$, $N_s=14$ will maximize $\mathbb{E}\{\eta\}$ for the \emph{user of interest} (and thus for all users with $P_e=0.01$). However, this will sacrifice the $\mathbb{E}\{\eta\}$ of users with high PERs (e.g. $P_e=0.3$ and $P_e=0.5$), and the PDR of other users are higher than $P_{th}=10^{-6}$.
On the other hand, choosing $N_s=37$ in Scenario II will guarantee the PDR to be smaller than $10^{-3}$ for all users, but this is achieved with very large margin for users with small values of PER. In fact, this is not desirable for these users as their throughputs are sacrificed.

\begin{table*} %[!t]
\ifCLASSOPTIONonecolumn
\renewcommand{\arraystretch}{.76}
\else
\renewcommand{\arraystretch}{1.35}
\fi
\caption{Optimum value of $N_s$, along with the arithmetic mean of users' $\mathbb{E}\{\eta\}$ and $P_d$ for broadcasting scenarios. In Scenario IV, the geometric means are also provided in parentheses. }
\ifCLASSOPTIONonecolumn
\vspace{-5mm}
\fi
\label{Table_Ns_eta_PDR}
\centering
\begin{tabular}{|l|ccc|ccc|}
\hline
& \multicolumn{3}{c}{$P_{th}=10^{-3}$} & \multicolumn{3}{|c|}{$P_{th}=10^{-6}$} \\
%\cline{2-7}
\hline
Scenario&$N_s$ &$\overline{\mathbb{E}\{\eta\}}$ & $\overline{P_d}$ &$N_s$ &$\overline{\mathbb{E}\{\eta\}}$ & $\overline{P_d}$ \\
\hline
\multicolumn{1}{|l|} {I}  & $11$ & $5.88\times10^5$&$0.135$ & $14$ & $5.97\times10^5$&$0.088$ \\
\multicolumn{1}{|l|} {II}  & $37$ & $5\times10^5$&$1\times10^{-4}$ & $52$ & $4.33\times10^5$&$9\times10^{-8}$ \\
\multicolumn{1}{|l|} {III} & $32$ & $5.26\times10^5$&$7\times10^{-4}$ & $48$ &  $4.49\times10^5$&$6\times10^{-7}$\\
\multicolumn{1}{|l|} {IV}  & $15$ & $5.98\times10^5$&$0.071~(9\times10^{-5})$ & $18$ & $5.94\times10^5$&$0.041~(4\times10^{-7})$ \\
\hline
\end{tabular}
\ifCLASSOPTIONonecolumn
\vspace{-7mm}
\fi
\end{table*}

Considering the results for Scenarios III and IV, it can be seen that we have moved from the extreme cases to more intermediate cases. In Scenario III, where we have employed the arithmetic mean of the performance metrics, it can be seen that still the performances of the users with higher PERs limit the performances of other users. This is due to the fact that their PDRs are at least $10^2$ times higher than those of other users (for $N_s=32$), and therefore they have the dominant effect on the arithmetic mean. This effect has become less dominant in Scenario IV by using the geometric mean of the PDR.

Another clear conclusion from the results in Table~\ref{Table_Ns_eta_PDR} (and also in Figs.~\ref{Fig_Eta_and_PDR_vs_Ns}(a) and \ref{Fig_Eta_and_PDR_vs_Ns}(b)) is the trade-off between the mean throughput and PDR. This trade-off is not only between the two performance metrics of one user, but also among the performance metrics of all users. Hence, the choice of the operating point (i.e. $N_s$, here) affects the performance of each user and also the overall performance of the system.

\ifCLASSOPTIONonecolumn
\vspace{-2mm}
\fi
\subsection{One-round SRLNC Scheme- Broadcasting with Variable $M$ and $q$} \label{Section_5_f}
%\ifCLASSOPTIONtwocolumn
%\vspace{-4mm}
%\fi
So far in this paper, we have considered the number of packets in a block $M$ and the field size $q$, to be fixed. However, in this section, we also take into account the effect of $M$ and $q$ on the optimization of the performance metrics, mean throughput and PDR. Considering variable $M$ and $q$, the packet length $l$ will also be variable, and as a result, the value of PER will vary depending on the number of bits in a packet. In order to form a unified framework for the comparison of different schemes against different values of $M$ and $q$, similar to~\cite{Lucani:INFOCOM:2009:RLN}, we consider a fixed bit error rate (BER), denoted by $P_{e_{bit}}$, and calculate the PER for every $M$ and $q$ by using $P_e = 1-(1-P_{e_{bit}})^l$. Here, we consider four different classes of users with BERs of $P_{e_{bit}}=[10^{-6}, 10^{-5}, 5\times10^{-5}, 10^{-4}]$, and assume that the normalized number of users having these BERs are $\bar{N}=[0.3, 0.4, 0.2, 0.1]$, respectively.

We focus only on the last two scenarios in Section~\ref{Section_4_b}, where the goal was to maximize the mean $\mathbb{E}\{\eta\}$ with a constraint on the arithmetic or geometric mean of PDR. We repeat the process in the previous subsection for different values of $M$ and $q$ and find the optimum transmission scheme among all feasible transmission schemes. The results are presented in Table~\ref{Table_Ns_M_q_eta_PDR}. Our first observation is the trade-off between the mean throughput and PDR. For instance, providing a lower arithmetic or geometric mean PDR of $P_{th}=10^{-6}$ leads to a smaller mean $\mathbb{E}\{\eta\}$ compared to the case with $P_{th}=10^{-3}$. It is also observed that there exists a direct relation between $M$ and the mean throughput values. However, $M$ cannot be chosen arbitrarily large as satisfying the required QoS as well as the delivery deadline limit its value.

\begin{table*} %[!t]
\ifCLASSOPTIONonecolumn
\renewcommand{\arraystretch}{.76}
\else
\renewcommand{\arraystretch}{1.35}
\fi
\caption{Optimum SRLNC design parameters, and the arithmetic mean of users' $\mathbb{E}\{\eta\}$ and $P_d$ for broadcasting Scenarios III and IV. In Scenario IV, the geometric means are also provided in parentheses.}
\ifCLASSOPTIONonecolumn
\vspace{-5mm}
\fi
\label{Table_Ns_M_q_eta_PDR}
\centering
\begin{tabular}{|l|ccc|ccc|}
\hline
& \multicolumn{3}{c}{$P_{th}=10^{-3}$} & \multicolumn{3}{|c|}{$P_{th}=10^{-6}$} \\
%\cline{2-7}
\hline
Scenario&$\{M,N_s,q\}$ &$\overline{\mathbb{E}\{\eta\}}$ & $\overline{P_d}$ &$\{M,N_s,q\}$ &$\overline{\mathbb{E}\{\eta\}}$ & $\overline{P_d}$ \\
\hline
\multicolumn{1}{|l|} {III}  & $\{44,158,8\}$ & $9.8\times10^5$&$9.6\times10^{-4}$ & $\{33,159,8\}$ & $7.4\times10^5$&$8.6\times10^{-7}$ \\
\multicolumn{1}{|l|} {IV}  & $\{142,156,4\}$ & $2.7\times10^6$&$0.173~(9.3\times10^{-4})$ & $\{133,155,8\}$ & $2.6\times10^6$&$0.149~(7.7\times10^{-7})$ \\
\hline
\end{tabular}
\ifCLASSOPTIONonecolumn
\vspace{-7mm}
\fi
\end{table*}

\begin{figure}%[!h]
\centering
%\vspace{-2mm}
\includegraphics[width=3in]{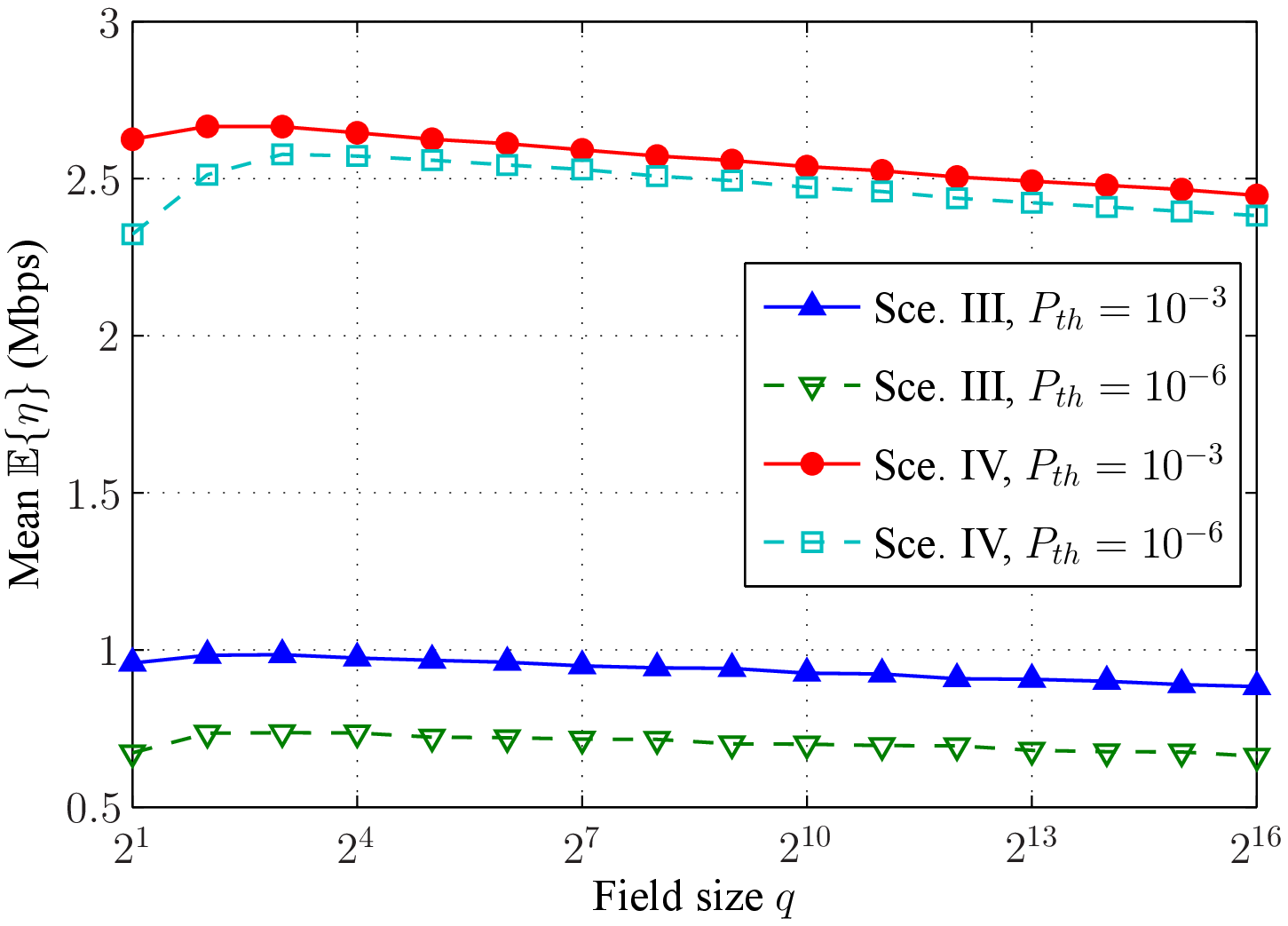}
\ifCLASSOPTIONonecolumn
\vspace{-7mm}
\fi
\caption{Optimum mean $\mathbb{E}\{\eta$\} versus field size $q$} \label{Fig_opt_Eta_vs_q}
\ifCLASSOPTIONonecolumn
\vspace{-8mm}
\fi
\end{figure}

The final observation is the effect of field size $q$, which is also highlighted in Fig.~\ref{Fig_opt_Eta_vs_q}. It can be seen that increasing the field size does not necessarily lead to a better performance. In fact, choosing very large $q$ in order to increase the probability of having linearly independent RLNC packets, not only imposes extra computational load for decoding of the packets to the users, but also results in extra overhead for sending the coding coefficients. This extra overhead degrades the mean throughput performance. Hence, instead of choosing very large values for $q$, a more  beneficial approach is to include it in the optimization of NC design parameters. This can potentially give rise to selecting practical values for $q$, as presented in Table~\ref{Table_Ns_M_q_eta_PDR}.

\subsection{Comparing One-round Schemes- Broadcasting with Variable $M$ and $q$} \label{Section_5_g}

\begin{figure*}%[!b]
\centering
\includegraphics[width=6.2in]{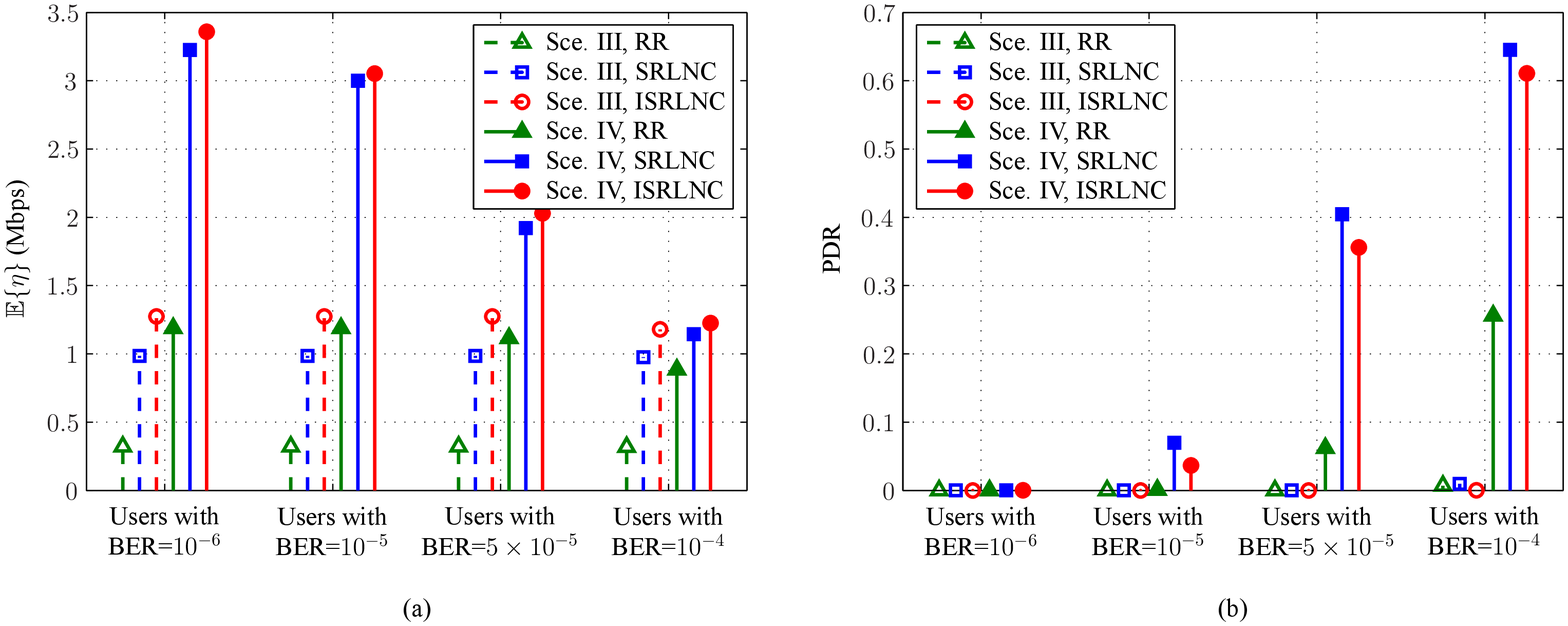}
\ifCLASSOPTIONonecolumn
\vspace{-5mm}
\fi
\caption{Comparison of the performance metrics for broadcasting scenarios. For Scenarios III and IV, the obtained design parameters are $\{M, N_s\}=\{14,154\}$ and $\{M, N_s\}=\{53,159\}$ for the RR scheme, and ${\{M, N_s\}=\{52,161\}}$ and ${\{M, N_s\}=\{151,161\}}$ for the ISRLNC scheme, respectively. The design variables for SRLNC scheme were presented in Table~\ref{Table_Ns_M_q_eta_PDR}.}
\label{Fig_Eta_vs_BER}
\ifCLASSOPTIONonecolumn
\vspace{-7mm}
\fi
\end{figure*}

In this subsection, we compare the broadcasting performance of SRLNC scheme with ISRLNC and RR schemes. We consider the same scenarios and parameters as in the previous subsection and provide the results for $P_{th}=10^{-3}$. We set the number of users to be $N=10$. The results showing $\mathbb{E}\{\eta\}$ and PDR for users with different BERs are depicted in Fig.~\ref{Fig_Eta_vs_BER}.

The first observation is that the performance of the SRLNC is very close to that of the ISRLNC scheme for both Scenarios III and IV. It is interesting that optimization of the design parameters of a feedback-less scheme results in performances close to those of an idealistic scheme with immediate feedbacks. 

By comparing the SRLNC and RR schemes, it can be observed that the throughput performance of SRLNC scheme is much better than that of the RR scheme for all classes of users in both scenarios. However, for Scenario IV, this comes at the cost of higher PDR in SRLNC compared to RR scheme for some classes of users. This is due to the fact the very small values of SRLNC's PDR for users with $P_{e_{bit}}=10^{-6}$ allow choosing larger values of $M$ in SRLNC compared to RR for maximizing the mean $\mathbb{E}\{\eta\}$, while satisfying the constraint on the geometric mean of PDR.

Having a closer look at the design parameters in Table~\ref{Table_Ns_M_q_eta_PDR} and also those used in Fig.~\ref{Fig_Eta_vs_BER}, it can be verified that the deadline requirement in each case is met with a very small or even no margin. In other words, the available time resources are not wasted as the transmission schemes are optimized for the best performance by exploiting them completely. This confirms the recommendation for the number of packets per block proposed in~\cite{Sorour:2010:ASMS:JCD}.

The final observation is the difference between the performances in Scenarios III and IV. It can be seen that in Scenario III, all users with different BERs achieve an almost similar mean throughput $\mathbb{E}\{\eta\}$ with a reasonably low PDR, whereas in Scenario IV, the users with lower BERs are serviced with much better throughputs and PDRs compared to those with higher BERs. This is the trade-off caused by the type of required QoS and is decided based on the application.

\ifCLASSOPTIONonecolumn
\vspace{-2mm}
\fi
\section{Conclusion}\label{Section_6}

In this paper, the problem of joint optimization of mean throughput and PDR in network coded (NC) systems for applications with strict delivery deadline requirements was studied. We employed RLNC and targeted satellite systems with TDD erasure channels and large RTTs. Here, we proposed a systematic framework to analytically study the mean throughout and PDR of users, as well as their interactions under various system parameters and settings. Using the proposed framework, the impact of feedback on the performance of the network coded systems under different RTTs and delivery deadline requirements was investigated. To this end, we compared the feedback-less schemes and schemes with feedback under the proposed unified framework, and observed that for systems with large RTTs, the feedback-less NC schemes provide better performances, in terms of the mean throughput and PDR.
Furthermore, we investigated the mean throughput and PDR of feedback-less NC schemes for a single-user case and highlighted the trade-off between these two performance metrics. Then, we extended our study to broadcasting to multiple users. We considered four different broadcasting scenarios and  for each of them obtained the best transmission scheme in terms of NC design parameters (i.e. the number of packets per block, the number of transmissions for each block and the field size). 
Finally, we compared the proposed feedback-less NC schemes with an ideal NC scheme, where immediate feedback about the reception status of each user was available at the sender. It was observed that by using the obtained optimized NC parameters for the feedback-less scheme, a performance very close to the ideal scheme 
can be achieved.

\appendices
\numberwithin{equation}{section}
\vspace{-8mm}
\section{Pareto Optimal Solutions} \label{Appendix_pareto}
This appendix provides details on solving the optimization problem in~\eqref{Eq_weighted_sum_two_round}, where $\mathbb{E}\{\eta\}$ and PDR are given in~\eqref{Eq_RLNC_two_round_mean_eta} and \eqref{Eq_RLNC_two_round_pdr}.
Having a closer look at these two equations, we can rewrite them in the following forms:
\vspace{-2mm}
\begin{align}
\mathbb{E}\{\eta\}&=A_0(N_s)+\sum_{j=1}^{M}A_j(N_s,N_j) \label{Eq_RLNC_two_round_mean_eta2} \\
&P_d=\sum_{j=1}^{M}B_j(N_s,N_j) \label{Eq_RLNC_two_round_pdr2}
\end{align}
where functions $A_0(\cdot)$, $A_j(\cdot)$ and $B_j(\cdot)$ can be obtained by replacing the throughputs and probabilities in~\eqref{Eq_RLNC_two_round_mean_eta} and \eqref{Eq_RLNC_two_round_pdr} with their corresponding values in~\eqref{Eq_P_two_round_RLNC} and~\eqref{Eq_eta_two_round_RLNC}. As a result,~\eqref{Eq_weighted_sum_two_round} can be rewritten as
\ifCLASSOPTIONonecolumn
\begin{align} \label{Eq_weighted_sum_two_round2}
\max_{N_s}\Big\{\lambda A_0(N_s)
+\sum_{j=1}^M\max_{N_j}\big\{\lambda A_i(N_s,N_j)-(1-\lambda)B_j(N_s,N_j)\big\}\Big\}
\end{align}
\else
\begin{align} \label{Eq_weighted_sum_two_round2}
\max_{N_s}\Big\{&\lambda A_0(N_s) \nonumber\\
&+\sum_{j=1}^M\max_{N_j}\big\{\lambda A_i(N_s,N_j)-(1-\lambda)B_j(N_s,N_j)\big\}\Big\}
\end{align}
\fi
where we have assumed a constant~$\lambda$ and feasible sets of $\{N_1,...,N_M,N_s\}$.

The solution to the above problem can be obtained by $M$ two-dimensional searches over values of $N_s$ and $N_j$. To compare the required computations of exhaustive and weighted sum methods for the two-round RLNC, we consider each design variable to take on average $k$ different values depending on $T_d$ and $T_{rt}$. In the exhaustive method, $\mathbb{E}\{\eta\}$ and $P_d$ should be calculated for $k^{M+1}$ feasible solutions to obtain the Pareto optimal points. However, in the weighted sum method described here, $Mk^2$ calculations of $\mathbb{E}\{\eta\}$ and $P_d$ are required to obtain one Pareto optimal point. Therefore, in order to have $v$ Pareto optimal points, then $vMk^2$ calculations of $\mathbb{E}\{\eta\}$ and $P_d$ are required. It can be observed that the required computational complexity of the weighted sum method is much lower compared to the exhaustive method for the practical values of $v$, $M$ and $k$.

It is worth noting that in order to obtain Pareto optimal points that are well distributed in the objective space, we chose $\lambda$ to be of the form $10^\theta$, where $\theta$ is chosen uniformly from interval $[-18,0]$.

For the SRLNC scheme, we will follow a similar approach. We use the exhaustive approach for the one-round scheme and the weighted sum method for the two-round scheme. For the weighted sum method in~\eqref{Eq_weighted_sum_two_round2}, the functions $A_0(\cdot)$, $A_j(\cdot)$ and $B_j(\cdot)$ can be obtained by using \eqref{Eq_P_two_round_SRLNC} and \eqref{Eq_eta_two_round_SRLNC}, and rewriting 
\eqref{Eq_SRLNC_two_round_mean_eta} and \eqref{Eq_SRLNC_two_round_pdr} in the forms of \eqref{Eq_RLNC_two_round_mean_eta2} and \eqref{Eq_RLNC_two_round_pdr2}.

\section{Performance Metrics for the Round Robin (RR) Scheme} \label{Appendix_RR}
As explained before, in the RR scheme, all packets in the block are transmitted $K$ times. Hence, the probability that a user successfully receives a packet after $K$ transmissions can be calculated as
\begin{align} \label{Eq_RR_packet_prob}
P_{PS|K}= \sum_{k=1}^K{{K}\choose{k}}(1-P_e)^kP_e^{K-k}
\end{align}
Then, the probability of receiving $m$ out of $M$ packets, after sending the block for $K$ times will be
\begin{align} \label{Eq_RR_packet_prob2}
P_{m|K}= {{M}\choose{m}}(P_{PS|K})^m(1-P_{PS|K})^{M-m}
\end{align}
In fact, this is the probability that the throughput takes the value of
\begin{align} \label{Eq_RR_eta}
\eta_{m|K}=\frac{mn}{KMT_{P_ u}+T_{rt}/2}
\end{align}

Thus, the performance metrics can be obtained as follow:
\begin{align}
\mathbb{E}\{\eta\}=\sum_{m=1}^MP_{m|K}\eta_{m|K} \label{Eq_RR_mean_eta}\\
P_d=\sum_{m=0}^{M-1}P_{m|K}\frac{M-m}{M} \label{Eq_RR_PDR}
\end{align}

\section{Performance Metrics for the Idealistic SRLNC (ISRLNC) Scheme} \label{Appendix_ISRLNC}
In this appendix, we obtain the performance metrics of the ISRLNC scheme for each user for various values of $N_s$. We start with $N_s>M$ and later will also provide the formulation for $N_s=M$.

As explained before, based on the immediate feedbacks from the users, if all users receive $M$ DOF after exactly $J$ transmissions ($M\leq J< N_s$), the sender stops the transmission. Otherwise, the transmission continues until the maximum number of transmissions $N_s$ is reached. In the first case, we denote the success sub-states by $\{S,J\}$ and write their probabilities as
\begin{align} \label{Eq_ISRLNC_prob}
P_{S,J}=\prod_{j=1}^N\sum_{k=M}^J{{J}\choose{k}}(1-P_{e_j})^k(P_{e_j})^{(J-k)}-\sum_{L=M}^{J-1}P_{S,L}
\end{align}
where the first and second terms in the right-hand side of this equation show the probabilities that all users receive $M$ DOF within $J$ and $J-1$ transmissions, respectively. For $J=M$, the second term should be set to zero. We note that these sub-states and probabilities are common among all users. 

In the second case, possible states for each user $i$ after $N_s$ transmissions are success sub-state $\{S,N_s,i\}$ and failure sub-states $\{F,N_s,i|m\}$, where $m<M$ is the number of uncoded packets received in the systematic phase of transmission. 
The probabilities for these states can be obtained as
\begin{align}
&P_{F,N_s,i|m}=\sum_{k=0}^{M-m-1}{{N_s-M}\choose{k}}(1-P_{e_i})^k(P_{e_i})^{N_s-M-k} \label{Eq_ISRLNC_prob4}\\
&P_{S,N_s,i}=1-\sum_{L=M}^{N_s-1}P_{S,L} -\sum_{m=0}^{M-1}P_{sys_i}(M,m)P_{F,N_s,i|m} \label{Eq_ISRLNC_prob3}
\end{align}
where~\eqref{Eq_ISRLNC_prob4} is given for cases in which the number of coded transmissions, i.e. $N_s-M$, is larger than or equal to the rDOF after the systematic phase, i.e. $M-m$. It can be easily inferred that for cases where $N_s-M<M-m$, this probability will be always equal to 1. In~\eqref{Eq_ISRLNC_prob3}, $P_{sys_i}(M,m)$, similar to~\eqref{Eq_P_sys}, is the probability that the $i$-th user receives $m$ out of $M$ uncoded packets in the systematic phase of transmission. Hence, the last term in the right-hand side of~\eqref{Eq_ISRLNC_prob3} shows the total probability that user $i$ receives lower than $M$ DOF after $N_s$ transmissions. The corresponding throughput values can thus be expressed as
\begin{align}
\eta= \left\{
\begin{array}{lcc}
\eta_{S,J}=\frac{Mn}{JT_{P_u}+T_{rt}/2} \label{Eq_ISRLNC_eta1}\\
\eta_{F,N_s,i|m}=\frac{mn}{N_sT_{P_u}+T_{rt}/2}\\ %\label{Eq_ISRLNC_eta2}
\eta_{S,N_s,i}=\frac{Mn}{N_sT_{P_u}+T_{rt}/2}
\end{array}\right.
\end{align}

Therefore, the $\mathbb{E}\{\eta_i\}$ and $P_{d_i}$ can be calculated through
\ifCLASSOPTIONonecolumn
\begin{align}
\mathbb{E}\{\eta_i\}=\left(\sum_{J=M}^{N_s-1}P_{S,J}\eta_{S,J}\right)&+P_{S,N_s,i}\eta_{S,N_s,i}
+\left(\sum_{m=1}^{M-1}P_{sys_i}(M,m)P_{F,N_s,i|m}\eta_{F,N_s,i|m}\right) \label{Eq_ISRLNC_ETA}\\
P_{d_i}=&\sum_{m=0}^{M-1}P_{sys_i}(M,m)P_{F,N_s,i|m}\frac{M-m}{M} \label{Eq_ISRLNC_PDR}
\end{align}
\else
\begin{align}
\mathbb{E}\{\eta_i\}=&\left(\sum_{J=M}^{N_s-1}P_{S,J}\eta_{S,J}\right)+P_{S,N_s,i}\eta_{S,N_s,i}\nonumber\\
+&\left(\sum_{m=1}^{M-1}P_{sys_i}(M,m)P_{F,N_s,i|m}\eta_{F,N_s,i|m}\right) \label{Eq_ISRLNC_ETA}\\
P_{d_i}=&\sum_{m=0}^{M-1}P_{sys_i}(M,m)P_{F,N_s,i|m}\frac{M-m}{M} \label{Eq_ISRLNC_PDR}
\end{align}
\fi

We should note that $\mathbb{E}\{\eta_i\}$ is affected by the number of users $N$, as increasing the value of $N$ decreases the probabilities in~\eqref{Eq_ISRLNC_prob}, and consequently reduces the mean throughput in~\eqref{Eq_ISRLNC_ETA}. 

For $N_s=M$,~\eqref{Eq_ISRLNC_eta1} to~\eqref{Eq_ISRLNC_PDR} are still valid, except for the first term in~\eqref{Eq_ISRLNC_ETA}, which is no more applicable; the value of $P_{F,N_s,i|m}$ will be always equal to $1$, and $P_{S,N_s,i}=(1-P_{e_i})^M$.

\ifCLASSOPTIONonecolumn
\vspace{-2mm}
\fi

%\ifCLASSOPTIONcaptionsoff
%  \newpage
%\fi

%\usepackage{setspace}
%\begin{spacing}{1.4}
%\newcommand{\BIBdecl}{\setlength{\itemsep}{-0.5 em}}
%\renewcommand{\arraystretch}{.7}
%\small 
\bibliographystyle{IEEEtran}

\ifCLASSOPTIONonecolumn
\linespread{1.43}
\fi
%\bibliography{Ref}

\begin{thebibliography}{10}
\providecommand{\url}[1]{#1}
\csname url@samestyle\endcsname
\providecommand{\newblock}{\relax}
\providecommand{\bibinfo}[2]{#2}
\providecommand{\BIBentrySTDinterwordspacing}{\spaceskip=0pt\relax}
\providecommand{\BIBentryALTinterwordstretchfactor}{4}
\providecommand{\BIBentryALTinterwordspacing}{\spaceskip=\fontdimen2\font plus
\BIBentryALTinterwordstretchfactor\fontdimen3\font minus
  \fontdimen4\font\relax}
\providecommand{\BIBforeignlanguage}[2]{{%
\expandafter\ifx\csname l@#1\endcsname\relax
\typeout{** WARNING: IEEEtran.bst: No hyphenation pattern has been}%
\typeout{** loaded for the language `#1'. Using the pattern for}%
\typeout{** the default language instead.}%
\else
\language=\csname l@#1\endcsname
\fi
#2}}
\providecommand{\BIBdecl}{\relax}
\BIBdecl

\bibitem{Shokrollahi:2006:IEEE-IT:Raptor}
A.~Shokrollahi, ``Raptor codes,'' \emph{IEEE Trans. Inform. Theory}, vol.~52,
  no.~6, pp. 2551--2567, June 2006.

\bibitem{Ho:IEEE-IT:2006:RLN}
T.~Ho, M.~M{\'e}dard, R.~Koetter, D.~Karger, M.~Effros, J.~Shi, and B.~Leong,
  ``A random linear network coding approach to multicast,'' \emph{IEEE Trans.
  Inform. Theory}, vol.~52, no.~10, pp. 4413--4430, Oct. 2006.

\bibitem{Chou:2003:Allerton:PNC}
P.~Chou, Y.~Wu, and K.~Jain, ``Practical network coding,'' in \emph{Proc.
  Allerton Conf. on Communication, Control, and Computing}, 2003.

\bibitem{Lun:2008:Phy-Com:CRC}
D.~S. Lun, M.~M{\'e}dard, R.~Koetter, and M.~Effros, ``On coding for reliable
  communication over packet networks,'' \emph{Phys. Commun.}, vol.~1, no.~1,
  pp. 3--20, Mar. 2008.

\bibitem{Eryilmaz:IEEE-IT:2008:DTG}
A.~Eryilmaz, A.~E. Ozdaglar, M.~M{\'e}dard, and E.~Ahmed, ``On the delay and
  throughput gains of coding in unreliable networks.'' \emph{IEEE Trans.
  Inform. Theory}, vol.~54, no.~12, pp. 5511--5524, 2008.

\bibitem{Lucani:INFOCOM:2009:RLN}
D.~E. Lucani, M.~Stojanovic, and M.~M{\'e}dard, ``Random linear network coding
  for time division duplexing: When to stop talking and start listening,'' in
  \emph{Proc. IEEE INFOCOM}, Rio de Janero, Brazil, Apr. 2009, pp. 1800--1808.

\bibitem{Lucani:NetCod:2009:BTD}
D.~E. Lucani, M.~M{\'e}dard, and M.~Stojanovic, ``Broadcasting in time-division
  duplexing: A random linear network coding approach,'' in \emph{Proc. IEEE
  NetCod}, Lausanne, Switzerland, June 2009, pp. 62--67.

\bibitem{Lucani:2009:GLOBECOM:RNC}
------, ``Random linear network coding for time-division duplexing: Field size
  considerations,'' in \emph{Proc. IEEE GLOBECOM}, Honolulu, Hawaii, USA,
  Nov./Dec. 2009, pp. 1--6.

\bibitem{Lucani:2010:ISIT:SysNC}
------, ``Systematic network coding for time-division duplexing,'' in
  \emph{Proc. IEEE ISIT}, June 2010, pp. 2403--2407.

\bibitem{Lucani:2012:IEEE-IT:NCD}
------, ``On coding for delay -- {N}etwork coding for time-division
  duplexing,'' \emph{IEEE Trans. Inform. Theory}, vol.~58, no.~4, pp.
  2330--2348, Apr. 2012.

\bibitem{Yazdi:2009:INFOCOM:ONC}
A.~Yazdi, S.~Sorour, S.~Valaee, and R.~Kim, ``Optimum network coding for delay
  sensitive applications in {W}i{M}{A}{X} unicast,'' in \emph{Proc. IEEE
  INFOCOM}, Apr. 2009, pp. 2576--2580.

\bibitem{Sorour:2009:PIMRC:NCARQ}
S.~Sorour and S.~Valaee, ``A network coded {A}{R}{Q} protocol for broadcast
  streaming over hybrid satellite systems,'' in \emph{Proc. IEEE PIMRC}, Sept.
  2009, pp. 1098--1102.

\bibitem{Sorour:2010:ASMS:JCD}
S.~Sorour, S.~Valaee, and N.~Alagha, ``Joint control of delay and packet drop
  rate in satellite systems using network coding,'' in \emph{Proc. Advanced
  Satellite Multimedia Systems Conf.}, Sept. 2010, pp. 46--53.

\bibitem{Swapna:2010:TDA}
B.~Swapna, A.~Eryilmaz, and N.~Shroff, ``Throughput-delay analysis of random
  linear network coding for wireless broadcasting,'' in \emph{Proc. IEEE
  NetCod}, June 2010, pp. 1--6.

\bibitem{Nistor:2011:IEEE-JSAC:ODD}
M.~Nistor, D.~E. Lucani, T.~Vinhoza, R.~A. Costa, and J.~Barros, ``On the delay
  distribution of random linear network coding,'' \emph{IEEE J. Sel. Areas
  Commun.}, vol.~29, no.~5, pp. 1084--1093, May 2011.

\bibitem{YANG:2012:ONLINE:ANC}
\BIBentryALTinterwordspacing
L.~Yang, Y.~Sagduyu, J.~Li, and J.~Zhang. (2012) Adaptive network coding for
  scheduling real-time traffic with hard deadlines. [Online]. Available:
  \url{http://arxiv.org/abs/1203.4008}
\BIBentrySTDinterwordspacing

\bibitem{Zeng:2012:ONLINE:JCD}
\BIBentryALTinterwordspacing
W.~Zeng, C.~Ng, and M.~Medard. (2012) Joint coding and scheduling optimization
  in wireless systems with varying delay sensitivities. [Online]. Available:
  \url{http://arxiv.org/abs/1202.0784}
\BIBentrySTDinterwordspacing

\bibitem{Tran:2012:ASM}
T.~Tran, H.~Li, W.~Lin, L.~Liu, and S.~Khan, ``Adaptive scheduling for
  multicasting hard deadline constrained prioritized data via network coding,''
  in \emph{Proc. IEEE Globcom}, Dec. 2012, pp. 1--5.

\bibitem{Goyal:2001:IEEE-SPM:MDC}
V.~Goyal, ``Multiple description coding: compression meets the network,''
  \emph{IEEE Signal Process. Mag.}, vol.~18, no.~5, pp. 74--93, Sept. 2001.

\bibitem{Sadeghi:EURASIP:2010:OAN}
P.~Sadeghi, R.~Shams, and D.~Traskov, ``An optimal adaptive network coding
  scheme for minimizing decoding delay in broadcast erasure channels,''
  \emph{EURASIP J. Wireless Commun. and Networking}, June 2010.

\bibitem{Keller:2008:NetCod:OBNC}
L.~Keller, E.~Drinea, and C.~Fragouli, ``Online broadcasting with network
  coding,'' in \emph{Proc. IEEE NetCod}, Jan. 2008, pp. 1--6.

\bibitem{Sadeghi:2012:ONLINE:IDRL}
\BIBentryALTinterwordspacing
P.~Sadeghi and M.~Yu. (2012) Instantly decodable versus random linear network
  coding: A comparative framework for throughput and decoding delay
  performance. [Online]. Available: \url{http://arxiv.org/abs/1208.2387}
\BIBentrySTDinterwordspacing

\bibitem{Sawaragi:1985:TMO}
Y.~Sawaragi, H.~Nakayama, and T.~Tanino, \emph{Theory of multiobjective
  optimization (vol. 176 of Mathematics in Science and Engineering)}. Orlando, FL, USA: Academic Press Inc., 1985.

\bibitem{Marler:2004:SMO}
R.~T. Marler and J.~S. Arora, ``{Survey of multi-objective optimization methods
  for engineering},'' \emph{Structural and Multidisciplinary Optimization},
  vol.~26, no.~6, pp. 369--395, Apr. 2004.

\bibitem{Ehrgott:2005:MCO}
M.~Ehrgott, \emph{Multicriteria optimization}. Berlin, Heidelberg: Springer-Verlag, 2005.

\end{thebibliography}


%\end{spacing}

\end{document}